\documentstyle{ar209}

\renewcommand{\evensidemargin}{0pc}
\renewcommand{\oddsidemargin}{0pc}
\begin{document}
\input psfig.sty  

\jname{Annu. Rev. Nucl. and Part. Science}
\jyear{2001}
\jvol{1}
\ARinfo{1056-8700/97/0610-00}

\def\ms{\hspace{1.5mm}}
\def\Athpi{A_{3\pi}^{\Delta R}}
\def\tvij{\tilde{v}_{ij}}
\def\tv{\tilde{v}}
\def\dvij{\delta v_{ij}}
\newcommand{\boldtau}{\mbox{\boldmath$\tau$}}
\newcommand{\boldsigma}{\mbox{\boldmath$\sigma$}}
\newcommand{\case}[2]{\textstyle{#1\over#2}}

\title{Quantum Monte Carlo Calculations of Light Nuclei}

\markboth{Steven C. Pieper and R. B. Wiringa}
{Exact Calculations of Light Nuclei}

\author{Steven C. Pieper and R. B. Wiringa
\affiliation{Physics Division, Argonne National Laboratory, Argonne, IL 60439;\\
email: spieper@anl.gov, wiringa@anl.gov}}


\begin{abstract}
\date \\
\\

Accurate quantum Monte Carlo calculations of ground and low-lying excited
states of light p-shell nuclei are now possible for realistic nuclear
Hamiltonians that fit nucleon-nucleon scattering data.
At present, results for more than 30 different $(J^{\pi};T)$ states, plus
isobaric analogs, in $A \leq 8$ nuclei have been obtained with an 
excellent reproduction of the experimental energy spectrum.
These microscopic calculations show that nuclear structure, including
both single-particle and clustering aspects, can be explained starting 
from elementary two- and three-nucleon interactions.
Various density and momentum distributions, electromagnetic form factors,
and spectroscopic factors have also been computed, as well as
electroweak capture reactions of astrophysical interest.\\
\\

With permission from the Annual Review of Nuclear and Particle
Science. Final version of this material is scheduled to appear in the Annual
Review of Nuclear and Particle Science Vol. 51, to be published in December
2001 by Annual Reviews, http://AnnualReviews.org.

\end{abstract}

\maketitle

\section{INTRODUCTION}

A major goal of nuclear physics is to understand the stability,
structure, and reactions of nuclei as a consequence of the interactions
between individual nucleons.
This is an extremely challenging many-body problem, exacerbated by the fact 
that we do not know in detail what the interactions are.
Quantum chromodynamics, the fundamental theory for strong interactions, is so 
difficult to solve in the nonperturbative regime of low-energy nuclear physics
that no one is currently able to quantitatively describe the interactions between two
nucleons from first principles.
However, we do have a tremendous amount of experimental data on nucleon-nucleon
($N\!N$) scattering and we can construct accurate representations of 
$N\!N$ interactions by means of two-body potentials.
These potentials are complicated, depending on the relative positions,
spins, isospins, and momenta of the nucleons, which makes the problem of
finding accurate quantum-mechanical solutions of many-nucleon bound and
scattering states a demanding task.
Nevertheless, a variety of precise calculations have been carried out for 
the three- and four-nucleon systems which reveal an additional complication:
two-nucleon forces alone are inadequate to quantitatively explain either the 
bound-state properties or a variety of $Nd$ and $Nt$ scattering data.
This is not surprising, because meson-exchange theory and the composite
nature of nucleons indicate that many-nucleon forces, at least three-nucleon
forces, are important.
However, these are even more difficult to construct from first principles.
In like manner, meson-exchange introduces two-body charge and current 
operators, and calculations of electroweak transitions show that these
are also necessary to achieve agreement with data.

Despite these difficulties, tremendous progress has been made in the past
decade, both in the characterization of nuclear forces and currents, and
in the development of accurate many-body techniques for evaluating them.
A multienergy partial-wave analyses of elastic $N\!N$ scattering data 
below $T_{lab} = 350$ MeV was produced by the Nijmegen group in 
1993~\cite{SKRS93}, which demonstrated that over 4300 data 
points could be fit with a $\chi^2$ per datum $\approx$ 1.
This was achieved by careful attention to the known long-range part of
the $N\!N$ interaction and a rigorous winnowing of inconsistent data.
Their work inspired the construction of a number of new $N\!N$ potential 
models which gave comparably good fits to the database.

Concurrently, several essentially exact methods have been developed for the
study of few-nucleon ($A$ = 3,4) systems with such realistic $N\!N$ potentials.
The first accurate, converged calculations of three-nucleon ($N\!N\!N$) bound 
states were obtained by configuration-space Faddeev methods in 
1985~\cite{CPFG85}, while the first accurate $^4$He bound states were 
obtained by the Green's function Monte Carlo method in 1987~\cite{C87}.
Other methods of comparable accuracy have been developed since 
then for both $A$ = 3,4 bound states and for $N\!N\!N$ scattering.
A comprehensive review of many aspects of the few-nucleon problem can be found
in the article by Carlson and Schiavilla~\cite{CS98}.
The result of all the progress in the last decade is that we have obtained an 
excellent understanding of few-nucleon systems, including both structure and 
reactions, although a few unanswered problems remain.

Until recently, nuclei larger than $A=4$ have been described within the 
framework of the nuclear shell model or mean-field pictures such as Skyrme 
models.
The classic description of p-shell ($5 \leq A \leq 16$)
nuclei was developed by Cohen and Kurath in the 1960s~\cite{CK65}.
In that work, a large number of experimental levels were fit in terms of a
relatively small number of two-body matrix elements, but no direct
connection was made to the underlying $N\!N$ force.
Beginning with the work of Kuo and Brown~\cite{KB66}, approximate methods 
for relating the shell-model matrix elements to the underlying forces were
developed and applied to even larger nuclei.  
However, in most calculations, only the valence nucleons, i.e., the 
nucleons in the last unfilled shell, are treated as active, while
the core is inert.
Recently the first "no-core" shell model calculations, in which all the 
nucleons are active, have been made in the p-shell with effective
interactions that have been derived by a G-matrix procedure from a realistic 
underlying $N\!N$ force~\cite{NVB00}.  
These calculations agree with the other exact methods for the $A=3,4$ 
nuclei, but fully-converged calculations for larger systems have not yet 
been obtained.

The present article focuses on recent developments in quantum Monte Carlo 
methods that are making the light p-shell nuclei accessible at a level of 
accuracy close to that obtained for the s-shell ($A \leq 4$) nuclei.
The quantum Monte Carlo methods include variational Monte Carlo (VMC) and
Green's function Monte Carlo (GFMC) methods.
The VMC is an approximate method~\cite{LAPS81,CPW83,W91,APW95} that is used 
as a starting point for the more accurate GFMC calculations.
Because of its relative simplicity, the VMC has also been used to make first
studies of various reactions, such as the nuclear response to electron
\cite{WS98,LWW99} or pion scattering~\cite{LW01}, and for electroweak capture 
reactions of astrophysical interest~\cite{NWS01,N01}.
The GFMC is exact in the sense that binding energies can be calculated with an 
accuracy of better than 2\%; it has been used to compute ground states and 
low-lying excitations of all the $A \leq 8$ nuclei~\cite{C87,PPCW95,PPCPW97,
WPCP00}.
These studies of light p-shell nuclei have shown that
nuclear structure can be predicted from the elementary nuclear forces.
As presently implemented, both the VMC and GFMC methods require significant
additional computer resources for every nucleon added, so they probably will
be limited to systems of $A \leq 12$ nuclei.
A new method, whose accuracy lies between that of the VMC and GFMC methods, is
auxiliary-field diffusion Monte Carlo (AFDMC); it has been implemented 
for large pure neutron systems~\cite{SF99} and with further development should
be capable of handling nuclei above $A=12$.

Most of the calculations discussed here have been made using Hamiltonians 
containing the Argonne $v_{18}$ $N\!N$ potential~\cite{WSS95} alone or with 
one of the Urbana~\cite{CPW83} or Illinois~\cite{PPWC01} series of $N\!N\!N$ 
potentials.
Argonne $v_{18}$ (AV18) is representative of the modern 
$N\!N$ potentials that give accurate fits to scattering data, while the 
Urbana and Illinois models are modern $N\!N\!N$ potentials based on 
meson-exchange interactions and fit to the binding energies of light nuclei.
These models are described in some detail in Section 2 to show the complexity
of modern nuclear forces, and the consequent challenge for many-body theory.
The VMC method is presented in Section 3; it starts with the construction of a 
variational wave function of specified angular momentum, parity and isospin,
$\Psi_V(J^{\pi};T)$, using products of two- and three-body correlation 
operators acting on a fully antisymmetrized set of 
one-body basis states coupled to specific quantum numbers.
Metropolis Monte Carlo integration~\cite{MR2T2} is used to evaluate 
$\langle \Psi_V | H | \Psi_V \rangle$, giving an upper bound to the 
energy of the state.
The GFMC method, described in Section 4, is a stochastic method that 
systematically improves on $\Psi_V$ by projecting out excited
state contamination using the Euclidean propagation
$\Psi(\tau) = \exp [ - ( H - E_0) \tau ] \Psi_V$.
In the limit $\tau \rightarrow \infty$, this leads to the exact 
$\langle | H | \rangle$.

Energy results for the ground states and many low-lying excited states of
the $A\leq8$ nuclei are presented in Section 5.
With the new Hamiltonians, the excitation spectra are reproduced very 
well, including the relative stability of different nuclei and the
splittings between excited states.
Energy differences between isobaric analog states and isospin-mixing matrix
elements have also been computed to study the charge-independence-breaking 
of the strong and electromagnetic forces.
A variety of density and momentum distributions, 
electromagnetic moments and form factors, transition densities, spectroscopic 
factors, and intrinsic deformation are presented in Section 6.
One of the chief benefits of this approach is that absolute normalizations
can be computed, and effective charges so commonly needed in traditional
shell-model calculations of electroweak transitions are not required.

While the quantum Monte Carlo methods have been used primarily for 
stationary states, it is also possible to study scattering states, as 
discussed in Section 7.
Some first calculations of astrophysical radiative capture reactions in the
p-shell are presented in Section 8.
Both of these applications are in their early stages, but promise to be
major fields of endeavor for future work.
In Section 9 we discuss neutron drops, systems of interacting neutrons 
confined in an artificial external well, that can serve as reference points 
for the development of Skyrme interactions for use in very large neutron-rich
nuclei.
Finally, in Section 10, we consider the outlook for future developments.

\section{HAMILTONIAN AND CURRENTS}

The Hamiltonian used in this review use includes nonrelativistic kinetic energy, 
the AV18 $N\!N$ potential~\cite{WSS95} and either the Urbana IX~\cite{PPCW95} 
or one of the new Illinois series of $N\!N\!$ potential~\cite{PPWC01}:
\begin{equation}
   H = \sum_{i} K_{i} + \sum_{i<j} v_{ij} + \sum_{i<j<k} V_{ijk} \ .
\label{eq:H}
\end{equation}
The kinetic energy operator has charge-independent (CI) and
charge-symmetry-breaking (CSB) components, the latter due to
the difference between proton and neutron masses,
\begin{equation}
   K_{i} = K^{CI}_{i} + K^{CSB}_{i} \equiv -\frac{\hbar^2}{4}
     (\frac{1}{m_{p}} + \frac{1}{m_{n}}) \nabla^{2}_{i}
    -\frac{\hbar^2}{4} (\frac{1}{m_{p}} - \frac{1}{m_{n}})\tau_{zi}
     \nabla^{2}_{i} ~,
\end{equation}
where $\boldtau_{i}$ is the isospin of nucleon $i$.

The AV18 potential is written as a sum of electromagnetic  and
one-pion-exchange terms and a shorter-range phenomenological part,
\begin{equation}
   v_{ij} = v^{\gamma}_{ij} + v^{\pi}_{ij} + v^{R}_{ij} \ .
\end{equation}
The electromagnetic terms include one- and two-photon-exchange Coulomb
interaction, vacuum polarization, Darwin-Foldy, and magnetic moment terms,
all with appropriate proton and neutron form factors.
The one-pion-exchange part of the potential includes the charge-dependent (CD)
terms due to the difference in neutral and charged pion masses.
It can be written as
\begin{equation}
   v^{\pi}_{ij} = f^{2} \left(\frac{m}{m_{s}}\right)^{2} \case{1}{3}
                 mc^{2} \left[ X_{ij} \boldtau_{i} \cdot \boldtau_{j}
                      + \tilde{X}_{ij} T_{ij} \right] \ ,
\label{eq:vpi}
\end{equation}
where $T_{ij} = 3\tau_{zi}\tau_{zj}-\boldtau_{i}\cdot\boldtau_{j}$ is the 
isotensor operator and
\begin{eqnarray}
   && X_{ij} = \case{1}{3} \left( X^{0}_{ij} + 2 X^{\pm}_{ij} \right) , \\
   && \tilde{X}_{ij} = \case{1}{3} \left( X^{0}_{ij} - X^{\pm}_{ij}
                                      \right) , \\
   && X^{m}_{ij} =  \left[ Y(mr_{ij}) \boldsigma_{i} \cdot \boldsigma_{j} +
                           T(mr_{ij}) S_{ij}  \right] \ .
\end{eqnarray}
Here $Y(x)$ and $T(x)$ are the normal Yukawa and tensor functions:
\begin{eqnarray}
Y(x) &=& \frac{e^{-x}}{x} \  \xi_Y(r) \ ,
\label{eq:yij}   \\
T(x) &=& \left( \frac{3}{x^2} + \frac{3}{x} + 1 \right) Y(x) \  \xi_T(r) \ .
\label{eq:tij}
\end{eqnarray}
with short-range cutoff functions $\xi_Y(r)$ and $\xi_T(r)$, and the 
$X^{\pm,0}$ are calculated with $m = m_{\pi^{\pm}}$ and $m_{\pi^{0}}$.
The coupling constant in Eq.(\ref{eq:vpi}) is $f^2 = 0.075$ and the scaling
mass $m_s = m_{\pi^{\pm}}$.

The remaining terms are of intermediate (two-pion-exchange) and short range,
with some 40 adjustable parameters.
The one-pion-exchange and the remaining phenomenological part of the potential
can be written as a sum of eighteen operators,
\begin{equation}
       v^{\pi}_{ij} + v^{R}_{ij} = \sum_{p=1,18} v_{p}(r_{ij}) O^{p}_{ij} \ .
\end{equation}
The first fourteen are CI operators,
\begin{eqnarray}
O^{p=1,14}_{ij} = [1, \boldsigma_{i}\cdot\boldsigma_{j}, S_{ij},
{\bf L\cdot S},{\bf L}^{2},{\bf L}^{2}(\boldsigma_{i}\cdot\boldsigma_{j}),
({\bf L\cdot S})^{2}]\otimes[1,\boldtau_{i}\cdot\boldtau_{j}] ~,
\label{eq:op14}
\end{eqnarray}
while the last four,
\begin{equation}
   O^{p=15,18}_{ij} = [1, \boldsigma_{i}\cdot\boldsigma_{j},
S_{ij}]\otimes T_{ij} \ ,  {\rm and} \ (\tau_{zi}+\tau_{zj}) \ ,
\end{equation}
are CD and CSB terms.
The potential was fit directly to the Nijmegen $N\!N$ scattering data
base~\cite{SKRS93} containing 1787 $pp$ and 2514 $np$ data in the range 
0--350 MeV, to the $nn$ scattering length, and to the deuteron binding energy,
with a $\chi^2$ per datum of 1.09.

While the CD and CSB terms are small, there is a clear need for their presence.
The Nijmegen group studied many $N\!N$ potentials from the 1980s and
before, and found that potentials fit to $np$ data in $T=1$ states did
not fit $pp$ data well even after allowing for standard Coulomb effects,
and vice versa~\cite{SS93-95}.
All five modern $N\!N$ potentials, Reid93, Nijm I and II~\cite{SKTS94},
CD Bonn~\cite{MSS96}, and AV18, which fit $pp$ and $np$
scattering data with a $\chi^2/$point near 1, have CD components.
The CSB term is required to fit the difference between the $pp$ and
$nn$ scattering lengths, and is consistent
with the mass difference between $^3$H and $^3$He, as shown below.
The identification of the proper CI, CD, and CSB $N\!N$ force components is 
important in setting the proper baseline for the introduction of $N\!N\!N$
forces, which are required to fit the $^3$H and $^3$He binding
energies as a primary constraint.

\begin{figure}
\centerline{\psfig{figure=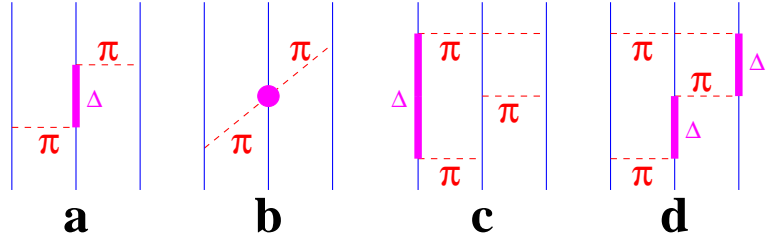,width=4.0in}}
\caption{Three-body force Feynman diagrams: a is the Fujita-Miyazawa or 
two-pion P-wave, b is two-pion S-wave, c and d are three-pion rings with 
one $\Delta$ in intermediate states.}
\label{fig:vijk}
\end{figure}

The Urbana series of three-nucleon potentials is written as a sum of
two-pion-exchange P-wave and remaining shorter-range phenomenological terms,
\begin{equation}
   V_{ijk} = V^{2\pi,P}_{ijk} + V^{R}_{ijk} ~.
\end{equation}
The structure of the two-pion P-wave exchange term with an intermediate
$\Delta$ excitation (Fig.~\ref{fig:vijk}a) 
was originally written down by Fujita and Miyazawa~\cite{FM57}; it
can be expressed simply as
\begin{equation}
   V^{2\pi,P}_{ijk} = \sum_{cyc} \left( A^P_{2\pi}
     \{X^{\pi}_{ij},X^{\pi}_{jk}\}
     \{\boldtau_{i}\cdot\boldtau_{j},\boldtau_{j}\cdot\boldtau_{k}\}
   + C^P_{2\pi} [X^{\pi}_{ij},X^{\pi}_{jk}]
     [\boldtau_{i}\cdot\boldtau_{j},\boldtau_{j}\cdot\boldtau_{k}] \right ) \ ,
\end{equation}
where $X^{\pi}_{ij}$ is constructed with an average pion mass,
$m_{\pi}=\case{1}{3}m_{\pi^0}+\case{2}{3}m_{\pi^{\pm}}$, and $\sum_{cyc}$
is a sum over the three cyclic exchanges of nucleons $i,j,k$.
For the Urbana models $C^P_{2\pi} = \case{1}{4}A^P_{2\pi}$, as in the original
Fujita-Miyazawa model~\cite{FM57}, while other potentials like the 
Tucson-Melbourne~\cite{TM} and Brazil~\cite{Brazil} models, have a ratio 
slightly larger than $\case{1}{4}$.
The shorter-range phenomenological term is given by
\begin{equation}
   V^{R}_{ijk} = \sum_{cyc} A_R T^2(m_{\pi}r_{ij}) T^2(m_{\pi}r_{jk}) \ .
\label{eq:vrijk}
\end{equation}
For the Urbana IX (UIX) model, the parameters were determined by fitting 
the binding energy of $^3$H and the density of nuclear matter in conjunction 
with AV18.

As shown below, while the combined AV18/UIX Hamiltonian reproduces the
binding energies of s-shell nuclei, it does not do so well for light 
p-shell nuclei.
Recently a new class of $N\!N\!N$ potentials, called the Illinois models, has 
been developed to address this problem~\cite{PPWC01}.  
These potentials contain the Urbana terms and two additional terms, resulting 
in a total of four coupling constants that can be adjusted to fit the data.  

One term, $V^{2\pi,S}_{ijk}$, is due to $\pi N$ S-wave scattering as 
illustrated in Fig.~\ref{fig:vijk}b. 
It has been included in a number of $N\!N\!N$ potentials like the 
Tucson-Melbourne~\cite{TM} and Brazil~\cite{Brazil} models.
The Illinois models use the form recommended in the latest Texas model~\cite{Texas}, 
where chiral symmetry is used to constrain the structure of the interaction.
However, in practice, this term is much smaller than the $V^{2\pi,P}_{ijk}$
contribution and behaves similarly in light nuclei, so it is difficult to
establish its strength independently just from calculations of energy levels.

A more important addition is a simplified form for three-pion rings
containing one or two deltas (Fig.~\ref{fig:vijk}c,d).  As discussed
in Ref.~\cite{PPWC01}, these diagrams result in a large number of terms,
the most important of which are used to construct the Illinois models:
\begin{equation}
V^{3\pi,\Delta R}_{ijk} = \Athpi \left[ \case{50}{3}S^I_{\tau}S^I_{\sigma}+
\case{26}{3}A^I_{\tau}A^I_{\sigma} \right] \ .
\end{equation}
Here the $S^I_x$ and $A^I_x$ are operators that are symmetric or antisymmetric
under any exchange of the three nucleons, and the subscript $x=\sigma$ or 
$\tau$ indicates that the operators act on, respectively, spin and space 
or just isospin degrees of freedom.

The $S^I_\tau$ is a projector onto isospin-$\case{3}{2}$ triples:
\begin{equation}
S^I_{\tau} = 2 + \case{2}{3}\left(\boldtau_i \cdot \boldtau_j 
+\boldtau_j \cdot \boldtau_k + \boldtau_k \cdot \boldtau_i \right) 
= 4 P_{T=3/2}  \ .  
\label{eq:sitau}
\end{equation}
To the extent isospin is conserved, there are no such triples in the s-shell 
nuclei, and so this term does not affect them.  It is also zero for $Nd$ 
scattering.  However, the $S^I_{\tau}S^I_{\sigma}$ term is 
attractive in all the p-shell nuclei studied.  The $A^I_\tau$ has the
same structure as the isospin part of anticommutator part of $V^{2\pi,P}$, 
but the $A^I_{\tau}A^I_{\sigma}$ term is repulsive in all nuclei studied so far.
In p-shell nuclei, the magnitude of the $A^I_{\tau}A^I_{\sigma}$ term is
smaller than that of the $S^I_{\tau}S^I_{\sigma}$ term, so the net effect
of the $V^{3\pi,\Delta R}_{ijk}$ is slight repulsion in s-shell nuclei and
larger attraction in p-shell nuclei.
The reader is referred to Ref.~\cite{PPWC01}, and its appendix, for
the complete structure of $V^{3\pi,\Delta R}_{ijk}$.

Relativistic corrections to the Hamiltonian of Eq.(\ref{eq:H}) have been
studied in three- and four-body nuclei~\cite{CPS93,FPCS95,FPA99}.  We only
briefly outline the results here.  If a square-root kinetic energy,
\begin{equation}
K^{rel}_i = \sqrt{ p^2_i + m^2_i } - m_i \ ,
\end{equation}
where $p_i$ is the momentum of nucleon $i$, is introduced into the
Hamiltonian, then the $N\!N$ potential must be readjusted to
refit the $N\!N$ data.  This can be done for AV18 with only small 
adjustments of the parameters, to produce a $v^{rel}$.  VMC calculations show 
that the resulting $\langle K^{rel}+v^{rel} \rangle \sim \langle K+v \rangle$
for $^3$H and $^4$He; thus the square-root kinetic energy
may be neglected in light nuclei.

A second relativistic correction, the so-called ``boost correction,''
arises from the fact that the two-nucleon interaction $v_{ij}$
depends both on the relative momentum ${\bf p} = ({\bf p}_i - {\bf p}_j)/2$
and the total momentum ${\bf P} = {\bf p}_i + {\bf p}_j $ of the 
interacting nucleons,
\begin{equation}
v_{ij}({\bf p},{\bf P}) = \tvij({\bf p}) + \dvij({\bf p},{\bf P}) \ ,
\label{eq:tvdv}
\end{equation}
where $\delta v_{ij}({\bf p},{\bf P}=0)=0$.  Fitting the two-body potential
to two-nucleon scattering data determines only $\tvij$.  The
$\dvij$ can be determined from $\tvij$; a suitable 
approximation~\cite{FPA99,FPF95,F75} is
\begin{equation}
\delta{v_{ij}}({\bf P})=-\frac{P^2}{8m^2}\tvij+\frac{1}{8m^2}\,\left[\ {\bf P}
\cdot{\bf r}\;{\bf P}\cdot\mbox{\boldmath$\nabla$}, \tvij\ \right] \ ,
\label{eq:friar}
\end{equation}
where only the first six terms (the static terms) of $\tvij$ are retained
and the gradient operators do not act on $\tvij$.  The $\dvij$
is meant to be used only in first-order perturbation in a 
non-relativistic wave function.  Its expectation values in light nuclei are 
nearly proportional to $\langle V^R_{ijk} \rangle$~\cite{PPWC01},
and thus one can consider that part of $A_R$, Eq.(\ref{eq:vrijk}), 
comes from $\dvij$.
This proportionality does not, however, extend to nuclear matter or pure
neutron systems.  There one must use a reduced $A_R^*$ that does
not contain the part ascribable to $\dvij$, and explicitly evaluate
$\langle \dvij \rangle$; doing so results in a softer equation of state
for dense matter.

\section{VARIATIONAL MONTE CARLO}

The variational method can be used to obtain approximate solutions to the
many-body Schr\"{o}dinger equation, $H\Psi = E\Psi$, for a wide range of
nuclear systems, including few-body nuclei, light closed-shell nuclei,
nuclear matter, and neutron stars~\cite{W93}.
A suitably parameterized wave function, $\Psi_V$, is used to calculate an
upper bound to the exact ground-state energy,
\begin{equation}
   E_V = \frac{\langle \Psi_V | H | \Psi_V \rangle}
              {\langle \Psi_V   |   \Psi_V \rangle} \geq E_0 \ .
\label{eq:expect}
\end{equation}
The parameters in $\Psi_V$ are varied to minimize $E_V$, and the lowest value
is taken as the approximate solution.
Upper bounds to excited states are also obtainable, if they have different
quantum numbers from the ground state, or from small-basis diagonalizations
if they have the same quantum numbers.
The corresponding $\Psi_V$ can be used to calculate other properties,
such as particle density or electromagnetic moments, or to
start a Green's function Monte Carlo calculation.
In this section we describe the {\it ansatz} for $\Psi_V$ for light
nuclei and briefly review how the expectation value
is evaluated and the parameters of $\Psi_V$ are fixed.

The best variational wave function has the form~\cite{APW95}
\begin{equation}
     |\Psi_V\rangle = \left[1 + \sum_{i<j<k}(U_{ijk}+U^{TNI}_{ijk})
                              + \sum_{i<j}U^{LS}_{ij} \right]
                      |\Psi_P\rangle \ ,
\label{eq:bestpsiv}
\end{equation}
where the pair wave function, $\Psi_P$, is given by
\begin{equation}
     |\Psi_P\rangle = \left[ {\cal S}\prod_{i<j}(1+U_{ij}) \right]
                      |\Psi_J\rangle \ .
\label{eq:psip}
\end{equation}
The $U_{ij}$, $U^{LS}_{ij}$, $U_{ijk}$, and $U^{TNI}_{ijk}$ are noncommuting
two- and three-nucleon correlation operators, and ${\cal S}$ is a
symmetrization operator.
The form of the antisymmetric Jastrow wave function,
$\Psi_J$, depends on the nuclear state under investigation.
For the s-shell nuclei the simple form 
\begin{equation}
     |\Psi_J\rangle = \left[ \prod_{i<j<k}f^c_{ijk} ({\bf r}_{ik},{\bf r}_{jk})
                             \prod_{i<j}f_c(r_{ij}) \right]
                     |\Phi_A(JMTT_{3})\rangle \ 
\end{equation}
is used.
Here $f_c(r_{ij})$ and $f^c_{ijk}$ are central (spin-isospin independent) two-
and three-body correlation functions and $\Phi_A$ is an antisymmetrized
spin-isospin state, e.g.,
\begin{eqnarray}
 &&  |\Phi_{3}(\case{1}{2} \case{1}{2} \case{1}{2} \case{1}{2}) \rangle
        = {\cal A} |\uparrow p \downarrow p \uparrow n \rangle \ , \\
 &&  |\Phi_{4}(0 0 0 0) \rangle
        = {\cal A} |\uparrow p \downarrow p \uparrow n \downarrow n \rangle \ ,
\end{eqnarray}
with ${\cal A}$ the antisymmetrization operator.

The two-body correlation operators $U_{ij}$ and $U^{LS}_{ij}$ are sums
of spin, isospin, tensor, and spin-orbit terms:
\begin{eqnarray}
   U_{ij} &=& \sum_{p=2,6} \left[ \prod_{k\not=i,j}f^p_{ijk}({\bf r}_{ik}
              ,{\bf r}_{jk}) \right] u_p(r_{ij}) O^p_{ij} \ , \\
   U^{LS}_{ij} &=& \sum_{p=7,8} \left[ \prod_{k\not=i,j}f^p_{ijk}({\bf r}_{ik}
              ,{\bf r}_{jk}) \right] u_p(r_{ij}) O^p_{ij} \ ,
\end{eqnarray}
where the $O^p_{ij}$ were introduced in Eq.(\ref{eq:op14}).
The spin-orbit correlations are only summed in Eq.(\ref{eq:bestpsiv})
because of the extra computational expense of evaluating powers of
${\bf L\cdot S}$ that would occur if it was inserted in the symmetrized
product of Eq.(\ref{eq:psip}).

\begin{figure}
\centerline{\psfig{figure=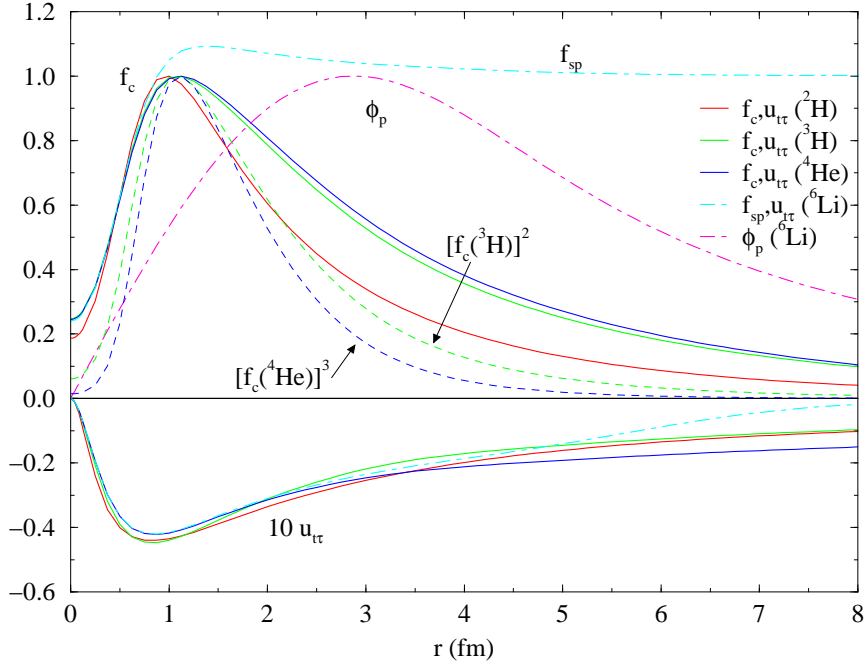,width=4.5in,angle=270}}
\caption{Correlation functions in light nuclei: the central pair correlation,
$f_c$, and tensor-isospin correlation, $u_{t\tau}$, for s-shell nuclei are 
shown with solid lines.  The dashed lines show $f_c^2(^3$H) and $f_c^3(^4$He) 
to illustrate the long-range separation behavior. The dot-dashed lines show 
the central correlation between s- and p-shell nucleons, $f_{sp}$, the
$u_{t\tau}$, and the one-body wave function, $\phi^{LS[n]}_p$, for $^6$Li.}
\label{fig:corr}
\end{figure}

The central $f_c(r)$ and noncentral $u_p(r)$ pair correlation
functions reflect the influence of the two-body potential at short distances,
while satisfying asymptotic boundary conditions of cluster separability.
Reasonable functions are generated by minimizing the two-body cluster
energy of a somewhat modified $N\!N$ interaction;
this results in a set of eight coupled, Schr\"{o}dinger-like, differential
equations corresponding to linear combinations of the first eight operators i
of $v_{ij}$, with a number of embedded variational parameters~\cite{W91}.
The $f_c(r)$ is small at short distances, to reduce the contribution
of the repulsive core of $v_{ij}$, and peaks at an intermediate distance
corresponding to the maximum attraction of $v_{ij}$, as illustrated in
Fig.~\ref{fig:corr} for several nuclei.
For the s-shell nuclei, $f_c(r)$ falls off at larger distances to keep the
system confined.
For example, in $^4$He, $[f_c(r)]^3 \sim {\rm {exp}}(-\kappa r)/r$ at large r,
as shown by the dashed line in Fig.~\ref{fig:corr},
where $\kappa$ corresponds to an $\alpha \rightarrow t+p$ separation energy.
The noncentral $u_p(r)$ are all relatively small; the most important
is the long-range tensor-isospin part $u_{t\tau}(r)$, also shown in
Fig.~\ref{fig:corr} for several nuclei, which is mainly
induced by the one-pion-exchange part of $v_{ij}$.

The $f^c_{ijk}$, $f^p_{ijk}$, and $U_{ijk}$ are three-nucleon correlations
also induced by $v_{ij}$~\cite{APW95}.
The $U^{TNI}_{ijk}$ are three-body correlations induced by the three-nucleon
interaction, which have the form suggested by perturbation theory:
\begin{equation}
     U^{TNI}_{ijk} = \sum_x \epsilon_x V_{ijk}(\tilde{ r}_{ij},
                     \tilde{r}_{jk}, \tilde{ r}_{ki}) \ ,
\label{eq:bestuijk}
\end{equation}
with $\tilde{r}=yr$, $y$ a scaling parameter, and $\epsilon_x$ a (small
negative) strength parameter.
A somewhat simpler wave function to evaluate than $\Psi_V$ is given by:
\begin{eqnarray}
|\Psi_{T}\rangle = \left[1 + \sum_{i<j<k}\tilde{U}^{TNI}_{ijk}\right]
|\Psi_P\rangle \ ,
\label{eq:psitgfmc}
\end{eqnarray}
where $\tilde{U}^{TNI}_{ijk}$ is a slightly truncated TNI correlation.
The $\Psi_T$ gives about $1-2$ MeV less binding than the full $\Psi_V$,
but costs less than half as much to construct, making it a more
efficient starting point for the GFMC calculation~\cite{PPCPW97}.

The Jastrow wave function for the light p-shell nuclei is more complicated, 
as a number of nucleons must be placed in the unfilled p-shell.
The $LS$ coupling scheme is used to obtain the desired $JM$ value of a given state,
as suggested by the shell-model studies of light p-shell nuclei~\cite{CK65}.
Different possible $LS$ combinations lead to multiple components in the
Jastrow wave function.
The possibility that the central correlations $f_{c}(r_{ij})$
could depend upon the shells (s or p) occupied by the
particles and on the $LS$ coupling is also allowed for:
\begin{eqnarray}
  |\Psi_J\rangle &=& {\cal A} \left\{
     \prod_{i<j<k}f^c_{ijk}
     \prod_{i<j \leq 4}f_{ss}(r_{ij})
     \prod_{k \leq 4 < l \leq A} f_{sp}(r_{kl}) \right. \\
  && \left.  \sum_{LS[n]} \Big( \beta_{LS[n]} \prod_{4 < l < m \leq A}
     f^{LS[n]}_{pp}(r_{lm})
     |\Phi_{A}(LS[n]JMTT_{3})_{1234:56\ldots A}\rangle \Big) \right\} \ .
     \nonumber 
\label{eq:jastrow}
\end{eqnarray}
The operator ${\cal A}$ indicates an antisymmetric sum over all possible
partitions of the $A$ particles into 4 s-shell and $(A-4)$ p-shell ones.
The central correlation $f_{ss}(r)$ is the $f_c(r)$ from the $^{4}$He
wave function. 
The $f_{sp}(r)$, shown in Fig.~\ref{fig:corr} for Li nuclei, is similar to 
the $f_c(r)$ at short range, but with a long-range tail going to unity; 
this helps the wave function factorize to a cluster structure like
$\alpha + d$ in $^6$Li or $\alpha + t$ in $^7$Li at large cluster separations.
The $f^{LS[n]}_{pp}(r)$ is similar to the deuteron (triton) $f_c(r)$ in the
case of $^6$Li ($^7$Li).

The $LS$ components of the single-particle wave function are given by:
\begin{eqnarray}
 &&  |\Phi_{A}(LS[n]JMTT_{3})_{1234:56\ldots A}\rangle =
     |\Phi_{\alpha}(0 0 0 0)_{1234}\rangle \prod_{4 < l\leq A}
     \phi^{LS[n]}_{p}(R_{\alpha l}) \\
 &&  \left\{ [ \prod_{4 < l\leq A} Y_{1m_l}(\Omega_{\alpha l}) ]_{LM_L[n]}
     \times [ \prod_{4 < l\leq A} \chi_{l}(\case{1}{2}m_s) ]_{SM_S}
     \right\}_{JM}
     \times [ \prod_{4 < l\leq A} \nu_{l}(\case{1}{2}t_3) ]_{TT_3}\rangle 
     \nonumber \ .
\label{eq:phi}
\end{eqnarray}
The $\phi^{LS[n]}_{p}(R_{\alpha k})$ are p-wave solutions of a particle
in an effective $\alpha + N$ potential and are functions of the distance
between the center of mass of the $\alpha$ core and nucleon $k$; they
may be different for different $LS[n]$ components.
A Woods-Saxon potential well,
\begin{equation}
     V_{\alpha N}(r) = V^{LS}_p [1+exp(\frac{r-R_p}{a_p})]^{-1} \ ,
\label{eq:spwell}
\end{equation}
where $V^{LS}_p$, $R_p$, and $a_p$ are variational parameters, and
a Coulomb term if appropriate is used.

For $A \geq 7$ nuclei, with three or more p-shell particles there are multiple 
ways of coupling the orbital angular momentum, $L$, spin, $S$, and isospin,
$T$, to obtain the total $(J^\pi;T)$ of the system.
The spatial permutation symmetry, denoted by the Young pattern $[n]$, is used
to enumerate the different possible terms, as discussed in Appendix 1C of
Ref.~\cite{BM69}.
The different possible contributions to $A$ = 6--8 nuclei are given in
Table~\ref{tab:perms}, with the corresponding spin states of highest
spatial symmetry for each nucleus.

\begin{table}
\def~{\hphantom{0}}
\caption{Permutation symmetry terms for $LS$-coupling in $A=6-8$ nuclei
and corresponding spin states.}
\label{tab:perms}
\begin{tabular}{lllll}
\toprule
  $A$  & $[n]  $ & $L    $ & $(T,S)                         $ 
& highest symmetry states                                   \\
\colrule
  $6$  & $[2]  $ & $0,2  $ & $(1,0) (0,1)                   $ 
& $^6$He$(0^+,2^+)$, $^6$Li$(1^+,2^+,3^+)$                  \\
       & $[11] $ & $1    $ & $(1,1) (0,0)                   $
& $^6$He$(1^+)$                                             \\
  $7$  & $[3]  $ & $1,3  $ & $(\case{1}{2},\case{1}{2})     $ 
& $^7$Li$(\case{1}{2}^-,\case{3}{2}^-,\case{5}{2}^-,\case{7}{2}^-)$ \\
       & $[21] $ & $1,2  $ & $(\case{3}{2},\case{1}{2})
                              (\case{1}{2},\case{3}{2})
                              (\case{1}{2},\case{1}{2})     $ 
& $^7$He$(\case{1}{2}^-,\case{3}{2}^-,\case{5}{2}^-)$ \\
       & $[111]$ & $0    $ & $(\case{3}{2},\case{3}{2})
                              (\case{1}{2},\case{1}{2})     $ \\
  $8$  & $[4]  $ & $0,2,4$ & $(0,0)                         $ 
& $^8$Be$(0^+,2^+,4^+)$                                     \\
       & $[31] $ & $1,2,3$ & $(1,1) (1,0) (0,1)             $
& $^8$Li$(0^+-4^+)$, $^8$Be$(1^+,3^+)$          \\
       & $[22] $ & $0,2  $ & $(2,0) (1,1) (0,2) (0,0)       $
& $^8$He$(0^+,2^+)$                                         \\
       & $[211]$ & $1    $ & $(2,1) (1,2) (1,1) (1,0) (0,1) $
& $^8$He$(1^+)$                                             \\
\botrule
\end{tabular}
\end{table}

After other parameters in the trial function have been optimized,  a
series of calculations are made in which the $\beta_{LS[n]}$ of Eq.(\ref{eq:jastrow})
may be different in the left- and right-hand-side wave functions to obtain the
diagonal and off-diagonal matrix elements of the Hamiltonian and the
corresponding normalizations and overlaps.
The resulting N$\times$N matrices are diagonalized to find the $\beta_{LS[n]}$
eigenvectors, using generalized eigenvalue routines because the correlated
$\Psi_V$ are not orthogonal.
This allows us to project out not only the ground states, but excited states
of the same $(J^\pi;T)$ quantum numbers.
For example, the ground states of $^6$Li, $^7$Li, and $^8$Li are obtained
from 3$\times$3, 5$\times$5, and 7$\times$7 diagonalizations, respectively.

The energy expectation value of Eq.(\ref{eq:expect}) is evaluated using Monte
Carlo integration.
A detailed technical description of the methods used can be found in
Refs.~\cite{W91,CW91}.
Monte Carlo sampling is done both in configuration space and in the
order of operators in the symmetrized product of Eq.(\ref{eq:psip})
by following a Metropolis random walk.
The expectation value for an operator $O$ is computed with the expression
\begin{equation}
  \langle O \rangle = \frac
  { \sum_{p,q} \int d{\bf R}
    \left[ \Psi_{p}^{\dagger}({\bf R}) O \Psi_{q}({\bf R}) /
           W_{pq}({\bf R}) \right] W_{pq}({\bf R}) }
  { \sum_{p,q} \int d{\bf R}
    \left[ \Psi_{p}^{\dagger}({\bf R})   \Psi_{q}({\bf R}) /
           W_{pq}({\bf R}) \right] W_{pq}({\bf R}) } \ ,
\end{equation}
where samples are drawn from a probability distribution, $W_{pq}({\bf R})$.
The subscripts $p$ and $q$ specify the order of operators in the left- and
right-hand-side wave functions, while the integration runs over the particle
coordinates ${\bf R}=({\bf r}_1,{\bf r}_2,\ldots,{\bf r}_A)$.
The probability distribution is constructed from the $\Psi_P$ of
Eq.(\ref{eq:psip}):
\begin{equation}
   W_{pq}({\bf R}) = | {\rm Re}( \langle \Psi_{P,p}^{\dagger}({\bf R})
                                       \Psi_{P,q}({\bf R}) \rangle ) | \ .
\label{eq:vmc:weight}
\end{equation}
This is much less expensive to compute than using the
full wave function of Eq.(\ref{eq:bestpsiv}) with its spin-orbit and
operator-dependent three-body correlations, but it typically
has a norm within 1--2\% of the full wave function.

Expectation values have a statistical error which can be estimated by the
standard deviation $\sigma$:
\begin{equation}
   \sigma = \left[ \frac{ \langle O^2 \rangle - \langle O \rangle ^2}
                        { N-1 } \right] ^{1/2} \ ,
\end{equation}
where $N$ is the number of statistically independent samples.
Block averaging schemes can also be used to estimate the autocorrelation
times and determine the statistical error.

The wave function $\Psi$ can be represented by an array of $2^A \times (^A_Z)$
complex numbers,
\begin{equation}
  \Psi({\bf R}) = \sum_{\alpha} \psi_{\alpha}({\bf R}) |\alpha\rangle \ ,
\label{eq:psivec}
\end{equation}
where the $\psi_{\alpha}({\bf R})$ are the complex coefficients of each state
$|\alpha\rangle$ with specific third components of spin and isospin.
This gives vectors with 96, 1280, 4480, and 14336 complex numbers
for $^4$He, $^6$Li, $^7$Li, and $^8$Li, respectively.
The spin, isospin, and tensor operators $O^{p=2,6}_{ij}$ contained in
the two-body correlation operator $U_{ij}$, and in the Hamiltonian are
sparse matrices in this basis.
For forces that are largely charge-independent, as is the case here,
this charge-conserving basis can be replaced with an isospin-conserving basis
that has $N(A,T) = 2^A \times I(A,T)$ components, where
\begin{equation}
   I(A,T) = \frac{2T+1}{\case{1}{2}A+T+1}
            \left(\begin{array}{c} A \\ \case{1}{2}A+T \end{array}\right) \ .
\label{eq:numiso}
\end{equation}
This reduces the number of vector elements to 32, 320, 1792, and 7168
for the cases given above --- a significant savings.
In practice, isospin operators are more expensive to evaluate
in this basis, but the overall savings in computation is still large.
Furthermore, if for even-$A$ nuclei the $M=0$ state is computed, only
half the spin components need to be evaluated; the other ones can be
found by time-reversal invariance.

Expectation values of the kinetic energy and spin-orbit potential require
the computation of first derivatives and diagonal second derivatives of the
wave function.
These are obtained by evaluating the wave function at $6A$ slightly shifted
positions of the coordinates ${\bf R}$ and taking finite differences,
as discussed in Ref.~\cite{W91}.
Potential terms quadratic in {\bf L} require mixed second derivatives, which
can be obtained by additional wave function evaluations and finite differences.

The rapid growth in the size of $\Psi$ as $A$ increases is the chief
limitation, both in computer time and memory requirements, on extending
the current method to larger nuclei.
The cost of an energy calculation for a given configuration also increases
as the number of pairs, $P = \case{1}{2}A(A-1)$, because of the number of
pair operations required to construct $\Psi$, and roughly as the number
of particles, $A$, due to the number of wave function evaluations required
to compute the kinetic energy by finite differences.
These factors are shown in Table~\ref{tab:scaling} for a number of nuclei,
with a final overall product showing the cost relative to that for an
$^8$Be calculation.
Some initial $A=9,10$ calculations have been made, but are at the limit of
current computer resources, so while $^{12}$C should be reached in a few
years, it may be the practical limit for this approach.
Calculations of nuclei like $^{16}$O or $^{40}$Ca, will require other
methods such as auxiliary-field diffusion Monte Carlo, which samples the spin and isospin components
of the wave function by introducing auxiliary fields, as the plain Monte 
Carlo samples the spatial part.
Also shown in the table is the cost for several pure neutron problems, such 
as an eight-body drop (with external confining potential) or 14 or 38 neutrons
in a box (with periodic boundary conditions) as neutron matter simulations.
The $^8$n drop has been calculated~\cite{PSCPPR96,SRP97} by VMC
and GFMC, and initial box simulations have been done for
14 neutrons by GFMC~\cite{carlson} and up to 54 neutrons by AFDMC~\cite{SF99}.

\begin{table}
\def~{\hphantom{0}}
\caption{Approximate scaling of VMC and GFMC calculations with system size.
The number of pairs is $P$, the state-vector size is the product of the number 
of spin states and number of isospin states, $I(A,T)$; as is discussed in the
text, these numbers may be halved for even-$A$ nuclei.
The last column is the relative difficulty scaled to $^8$Be, assuming
$M=0$ states are computed for even-$A$ nuclei.}
\label{tab:scaling}
\begin{tabular}{lrrll}
\toprule
         & $A$ & $P$ & $2^A\times I(A,T)$ & $\prod / \prod(^8$Be) \\
\colrule
   $^3$H   &  3  &   3 &     8$\times$2     &   0.0004    \\
   $^4$He  &  4  &   6 &    16$\times$2     &   0.001     \\
   $^5$He  &  5  &  10 &    32$\times$5     &   0.020     \\
   $^6$Li  &  6  &  15 &    64$\times$5     &   0.036     \\
   $^7$Li  &  7  &  21 &   128$\times$14    &   0.66      \\
   $^8$Be  &  8  &  28 &   256$\times$14    &   1.        \\
   $^9$Be  &  9  &  36 &   512$\times$42    &  17.        \\
 $^{10}$B  & 10  &  45 &  1024$\times$42    &  24.        \\
 $^{12}$C  & 12  &  66 &  4096$\times$132   & 530.        \\
 $^{16}$O  & 16  & 120 & 65536$\times$1430  &  $2 \times 10^5$  \\
 $^{40}$Ca & 40  & 780 & 7$\times 10^{21}$  &  $3 \times 10^{20}$  \\
\colrule
   $^8$n   &  8  &  28 &   256$\times$1     &   0.071     \\
 $^{14}$n  & 14  &  91 & 16384$\times$1     &  26.        \\
 $^{38}$n  & 38  & 703 & 3$\times 10^{11}$  &  $9 \times 10^9$ \\
\botrule
\end{tabular}
\end{table}

A major problem arises in minimizing the variational energy for p-shell
nuclei using the above wave functions: there is no variational minimum that
gives reasonable rms radii.
For example, the variational energy for $^6$Li is slightly more bound
than for $^4$He, but is not more bound than for separated $^4$He and $^2$H
nuclei, so the wave function is not stable against breakup into $\alpha + d$
subclusters.
Consequently, the energy can be lowered toward the sum of $^4$He and $^2$H
energies by making the wave function more and more diffuse.
Such a diffuse wave function would not be useful for computing other nuclear
properties, or as a starting point for the GFMC calculation,
so the search for variational parameters is constrained by requiring the
resulting point proton rms radius, $r_p$, to be close to the experimental
values for $^6$Li and $^7$Li ground states.
For other $A$ = 6--8 ground states, and all the excited states,
the trial functions contain minimal changes to the $^6$Li and $^7$Li wave
functions, with the added requirement that excited states should not have
smaller radii than the ground states.
The final step is always the diagonalization of the Hamiltonian in the
$\beta_{LS[n]}$ mixing parameters.

\section{GREEN'S FUNCTION MONTE CARLO}

The GFMC method~\cite{C87,C88} projects out the exact lowest-energy state,
$\Psi_{0}$, for a given set of quantum numbers, using
$\Psi_0 = \lim_{\tau \rightarrow \infty} \exp [ - ( H - E_0) \tau ] \Psi_T$,
where $\Psi_{T}$ is an initial trial function.
If the maximum $\tau$ actually used is large enough,
the eigenvalue $E_{0}$ is calculated exactly while other expectation values
are generally calculated neglecting terms of order $|\Psi_{0}-\Psi_{T}|^{2}$
and higher~\cite{PPCPW97}.
In contrast, the error in the variational energy, $E_{V}$, is of order
$|\Psi_{0}-\Psi_{T}|^{2}$, and other expectation values calculated with
$\Psi_{T}$ have errors of order $|\Psi_{0}-\Psi_{T}|$.
In the following we present a brief overview of modern nuclear GFMC
methods; much more detail may be found in Refs.~\cite{PPCPW97,WPCP00}.

We start with the $\Psi_{T}$ of Eq.(\ref{eq:psitgfmc}) and define the propagated
wave function $\Psi(\tau)$
\begin{eqnarray}
   \Psi(\tau) = e^{-({H}-E_{0})\tau} \Psi_{T}
              = \left[e^{-({H}-E_{0})\triangle\tau}\right]^{n} \Psi_{T} \ ,
\end{eqnarray}
where we have introduced a small time step, $\tau=n\triangle\tau$;
obviously $\Psi(\tau=0) =  \Psi_{T}$ and
$\Psi(\tau \rightarrow \infty) = \Psi_{0}$.
The $\Psi (\tau)$ is represented by a vector function of $\bf R$ using
Eq.(\ref{eq:psivec}), and the Green's function,
$G_{\alpha\beta}({\bf R},{\bf R}^{\prime})$ is a matrix function of
$\bf R$ and ${\bf R}^{\prime}$ in spin-isospin space, defined as
\begin{equation}
G_{\alpha\beta}({\bf R},{\bf R}^{\prime})= \langle {\bf
R},\alpha|e^{-({H}-E_{0})\triangle\tau}|{\bf R}^{\prime},\beta\rangle \ .
\label{eq:gfunction}
\end{equation}
It is calculated with leading errors of order $(\triangle\tau)^{3}$.
Omitting the spin-isospin indices $\alpha$, $\beta$ for brevity,
$\Psi({\bf R}_{n},\tau)$ is given by
\begin{equation}
\Psi({\bf R}_{n},\tau) = \int G({\bf R}_{n},{\bf R}_{n-1})\cdots G({\bf
R}_{1},{\bf R}_{0})\Psi_{T}({\bf R}_{0})d{\bf R}_{n-1}\cdots d{\bf R}_{1}d{\bf
R}_{0} \ ,
\label{eq:gfmcpsi}
\end{equation}
with the integral being evaluated stochastically.

The short-time propagator should allow as large a time step $\triangle\tau$ as
possible, because the total computational time for propagation is proportional
to $1/\triangle\tau$.
Earlier calculations \cite{PPCW95,C87,C88} used the propagator obtained from
the Feynman formulae in which the kinetic and potential energy terms
of $H$ are separately exponentiated.
The main error in this approximation to $G_{\alpha,\beta}$ comes from terms in
$e^{-{H}\triangle\tau}$ having multiple $v_{ij}$, like
$v_{ij}Kv_{ij}(\triangle\tau)^{3}$, where $K$ is the kinetic energy,
which can become large when particles $i$ and $j$ are very close
due to the large repulsive core in $v_{ij}$.
This requires a rather small $\triangle\tau \sim 0.1$
GeV$^{-1}$.

It has been found in studies of bulk helium atoms~\cite{C95}
that including the exact two-body propagator allows much larger time steps.
This short-time propagator is
\begin{eqnarray}
G_{\alpha\beta}({\bf R},{\bf R}^{\prime}) &=& G_{0}({\bf R},{\bf
R}^{\prime})\langle\alpha|\left[{\cal S}\prod_{i<j}\frac{g_{ij}({\bf
r}_{ij},{\bf r}_{ij}^{\prime})}{g_{0,ij}({\bf r}_{ij},{\bf r}_{ij}^{\prime})}
\right] |\beta\rangle \ ,
\end{eqnarray}
where
\begin{eqnarray}
G_{0}({\bf R},{\bf R}^{\prime}) &=& \langle {\bf R}|e^{-{K}\triangle\tau}|{\bf
R}^{\prime}\rangle =  \left[ \sqrt{\frac{m}
{2\pi\hbar^{2} \triangle\tau}}\, \right]^{3A}\exp\left[\frac{-({\bf R}-{\bf
R}^{\prime})^2}{2\hbar^{2}\triangle\tau/m}\right] \ ,
\label{eq:propagator2}
\end{eqnarray}
$g_{ij}$ is the exact two-body propagator,
\begin{eqnarray}
g_{ij}({\bf r}_{ij},{\bf r}_{ij}^{\prime}) = \langle{\bf
r}_{ij}|e^{-H_{ij}\triangle\tau}|{\bf r}_{ij}^{\prime}\rangle \ ,
\label{eq:gij}
\end{eqnarray}
and $g_{0,ij}$ is the free two-body propagator.
All terms containing any number of the same $v_{ij}$ and $K$ are treated
exactly in this propagator, as we have included the imaginary-time equivalent
of the full two-body scattering amplitude.
It still has errors of order $(\triangle\tau)^{3}$, however they are from
commutators of terms like $v_{ij}Kv_{ik}(\triangle\tau)^{3}$ which become
large only when both pairs $ij$ and $ik$ are close.
Because this is a rare occurrence, a five times larger time step,
$\triangle\tau \sim 0.5$ GeV$^{-1}$, can be used~\cite{PPCPW97}.

Including the three-body forces and the $E_{0}$ in
Eq.(\ref{eq:gfunction}), the complete propagator is given by
\begin{eqnarray}
G_{\alpha\beta}({\bf R},{\bf R}^{\prime})& = &e^{E_{o}\triangle\tau}G_{0}({\bf
R},{\bf R}^{\prime})\exp[{- \sum (V^{R}_{ijk}({\bf R})+ V^{R}_{ijk}({\bf
R^{\prime}}))\frac{\triangle\tau}{2}}] \nonumber \\
& & \langle\alpha|I_{3}({\bf R})|\gamma\rangle\langle\gamma|\left[{\cal
S}\prod_{i<j}\frac{g_{ij}({\bf r}_{ij},{\bf r}_{ij}^{\prime})}
{g_{0,ij}({\bf r}_{ij},{\bf r}_{ij}^{\prime})} \right] 
|\delta\rangle\langle\delta|I_{3}({\bf
R}^{\prime})|\beta\rangle~,
\label{eq:fullprop}
\end{eqnarray}
where
\begin{eqnarray}
I_{3}({\bf R}) = \left[1 - \frac{\triangle\tau}{2}\sum V^{\pi}_{ijk}({\bf
R})\right]~.
\end{eqnarray}
and $V^{\pi}_{ijk}=V^{2\pi}_{ijk}+V^{3\pi}_{ijk}$ represents, in general,  all
non-central $V_{ijk}$ terms.
With the exponential of $V^{\pi}_{ijk}$ expanded to first order in
$\triangle\tau$, there are additional error terms of the form
$V^{\pi}_{ijk}V^{\pi}_{i'j'k'}(\triangle\tau)^{2}$.
However, they have negligible effect because $V^{\pi}_{ijk}$ has a magnitude
of only a few MeV.

Quantities of interest are evaluated in terms of a ``mixed'' expectation value
between $\Psi_T$ and $\Psi(\tau)$:
\begin{eqnarray}
\langle O \rangle_{\rm Mixed} & = & \frac{\langle \Psi_{T} | O |
\Psi(\tau)\rangle}{\langle \Psi_{T} | \Psi(\tau)\rangle} \nonumber \\
& = & \frac{ \int d {\bf P}_n
\Psi_{T}^{\dagger}({\bf R}_{n}) O G({\bf R}_{n},{\bf
R}_{n-1})\cdots G({\bf R}_{1},{\bf R}_{0})\Psi_{T}({\bf R}_{0})}
{\int d{\bf P}_{n} \Psi_{T}^{\dagger}
({\bf R}_{n})G({\bf R}_{n},{\bf R}_{n-1}) \cdots G({\bf
R}_{1},{\bf R}_{0})\Psi_{T}({\bf R}_{0})}~,
\label{eq:expectation}
\end{eqnarray}
where ${\bf P}_{n} = {\bf R}_{0},{\bf R}_{1},\cdots,{\bf R}_{n}$ denotes the
`path', and
$d{\bf P}_{n} = d{\bf R}_{0} d{\bf R}_{1}\cdots d{\bf R}_{n}$ with
the integral over the paths being carried out stochastically.
The desired expectation values would, of course, have $\Psi(\tau)$ on both
sides; by writing
$\Psi(\tau) = \Psi_{T} + \delta\Psi(\tau)$ 
and neglecting terms of order $[\delta\Psi(\tau)]^2$, we obtain the approximate
expression
\begin{eqnarray}
\langle O (\tau)\rangle =
\frac{\langle\Psi(\tau)| O |\Psi(\tau)\rangle}
{\langle\Psi(\tau)|\Psi(\tau)\rangle}
\approx \langle O (\tau)\rangle_{\rm Mixed}
     + [\langle O (\tau)\rangle_{\rm Mixed} - \langle O \rangle_T] ~,
\label{eq:pc_gfmc}
\end{eqnarray}
where $\langle O \rangle_T$ is the variational expectation value.
More accurate evaluations of $\langle O (\tau)\rangle$ are possible~\cite{K67},
essentially by measuring the observable at the mid-point of the path.
However, such estimates require a propagation twice as long as the mixed
estimate and require separate propagations for every $\langle O \rangle$
to be evaluated.  The nuclear calculations published to date use the
approximation of Eq.(\ref{eq:pc_gfmc}).

A special case is the expectation value of the Hamiltonian.
The $\langle{H}(\tau)\rangle_{\rm Mixed}$ can be re-expressed as~\cite{CK79}
\begin{eqnarray}
\langle{H}(\tau)\rangle_{\rm Mixed} = \frac{\langle \Psi_{T} |
e^{-({H}-E_{0})\tau /2}{H} e^{-({H}-E_{0})\tau /2} |
\Psi_{T}\rangle}{\langle \Psi_{T} |e^{-({H}-E_{0})\tau /2}
e^{-({H}-E_{0})\tau /2}| \Psi_{T}\rangle} \geq E_{0}~,
\end{eqnarray}
since the propagator $\exp [ - (H - E_0) \tau ] $ commutes with the
Hamiltonian.
Thus $\langle{H}(\tau)\rangle_{\rm Mixed}$ approaches $E_{0}$ in the limit
$\tau\rightarrow\infty$, and the expectation value obeys the variational
principle for all $\tau$.

The AV18 interaction contains terms that are quadratic in orbital angular
momentum, $L$.  These terms are, in essence, state- and position-dependent
modifications of the mass of the nucleons.  If they are included in
the calculation of $g_{ij}$, then the ratio $g_{ij}/g_{0,ij}$ in
Eq.(\ref{eq:propagator2}) becomes unbounded for large 
$|{\bf r}_{ij}|$ or $|{\bf r}_{ij}^{\prime}|$ and the Monte Carlo statistical
error will also be unbounded.  
Hence the GFMC propagator is constructed with a simpler isoscalar
interaction, $H^{\prime}$, with a $v^{\prime}_{ij}$
that has only eight operator terms,
$[1, \boldsigma_{i}\cdot\boldsigma_{j}, S_{ij}, {\bf L\cdot S}]\otimes
[1, \boldtau_{i}\cdot\boldtau_{j}]$, chosen such that it equals the CI
part of the full interaction in all S- and P-waves and in the deuteron.
The $v^{\prime}_{ij}$ is a little more attractive than $v_{ij}$, so
a $V^{\prime}_{ijk}$ adjusted so that
$\langle H^{\prime} \rangle \approx \langle H \rangle$ is also used.
This ensures the GFMC algorithm will not propagate to excessively large
densities due to overbinding.
Consequently, the upper bound property applies to
$\langle{H}^{\prime}(\tau)\rangle$, and $\langle H-H^{\prime} \rangle$ must
be evaluated perturbatively.

Another complication that arises in the GFMC algorithm is the ``fermion sign
problem''.
This arises from the stochastic evaluation of the matrix elements
in Eq.(\ref{eq:expectation}).  The $G({\bf R}_{i},{\bf R}_{i-1})$ is
a local operator and can mix in the boson solution.  This has a (much) lower
energy than the fermion solution and thus is exponentially amplified
in subsequent propagations.  In the final integration with the antisymmetric
$\Psi_T$, the desired fermionic part is projected out, but in the
presence of large statistical errors that grow exponentially with $\tau$.
Because the number of pairs that can be exchanged grows with $A$, the
sign problem also grows exponentially with increasing $A$.  
For $A{\geq}8$, the errors grow so fast that convergence in $\tau$ cannot
be achieved.

For simple scalar wave functions, the fermion sign problem can be controlled
by not allowing the propagation to move across a node of the wave function.
Such ``fixed-node'' GFMC provides an approximate solution which is the
best possible variational wave function with the same nodal structure 
as $\Psi_T$.
However, a more complicated
solution is necessary for the spin- and isospin-dependent wave functions
of nuclei.  In the last few years a suitable ``constrained path''
approximation has been developed and extensively tested, first for condensed
matter systems~\cite{ZCG95} and more recently for nuclei~\cite{WPCP00}.
The basic idea of the constrained-path method is to discard those 
configurations that, in future generations, will contribute only noise to 
expectation values.  
If the exact ground state $| \Psi_0 \rangle$ were known,
any configuration for which
\begin{equation}
\Psi({\bf P}_n)^\dagger \Psi_0({\bf R}_n) = 0 \ ,
\label{eq:gfmc:const_config}
\end{equation}
where a sum over spin-isospin states is implied, could be discarded.
Here ${\bf P}_n$ designates the complete path, $({\bf R}_0, ..., {\bf R}_n)$
that has led to ${\bf R}_n$.
The sum of these discarded
configurations can be written as a state $| \Psi_d \rangle$,
which obviously has zero overlap with the ground state.
The $\Psi_d$ contains only excited states and should decay away as 
$\tau \rightarrow \infty$, thus discarding it is justified. 
Of course the exact $\Psi_0$ is not known, and so configurations
are discarded with a probability such that the average overlap
with the trial wave function, 
\begin{equation}
\langle \Psi_d | \Psi_T \rangle = 0 \ .
\end{equation}
Many tests of this procedure have been made~\cite{WPCP00} and it usually gives
results that are consistent with unconstrained propagation, within statistical errors.
However a few cases in which the constrained propagation converges to the
wrong energy (either above or below the correct energy) have been found.
Therefore a small number, $n_u=10$ to 20, unconstrained steps are made
before evaluating expectation values.  These few unconstrained steps,
out of typically 400 total steps,
appear to be enough to damp out errors introduced by the constraint,
but do not greatly increase the statistical error.

\begin{figure}
\centerline{\psfig{figure=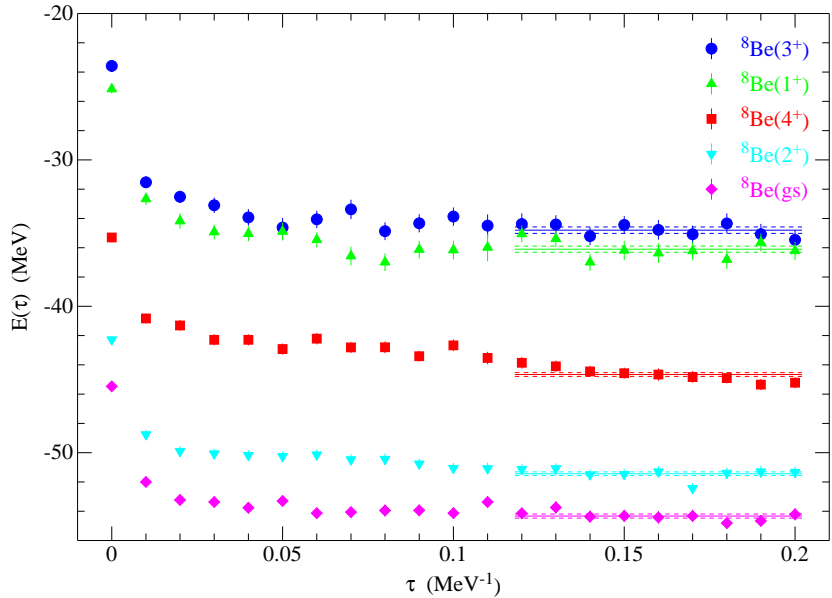,width=4.5in,angle=270}}
\caption{GFMC $E(\tau)$ using AV18/UIX for various states of $^8$Be.
Averages with their standard errors are shown by the solid and dashed lines.}
\label{fig:8be-e_of_tau}
\end{figure}

Figure \ref{fig:8be-e_of_tau} shows the progress with increasing $\tau$
of typical constrained GFMC calculations, in this case for various
states of $^8$Be.  The values shown at $\tau=0$ are the VMC
values using $\Psi_T$; VMC values with the best known $\Psi_V$ are
about 1.5 MeV below these.  The GFMC very rapidly makes a large
improvement on these energies; by $\tau=0.01$~MeV$^{-1}$, the
$\Psi_T$ energies have been reduced by 6.5 to 8~MeV.  This rapid
improvement corresponds to the removal of small admixtures of 
states with excitation energies $\sim~1$~GeV from $\Psi_T$.
Averages over typically the last nine $\tau$ values are
used as the GFMC energy.  The standard deviation, computed using
block averaging, of all of the individual energies for these $\tau$ values
is used to compute the corresponding statistical error.  
The solid lines show these averages; the corresponding dashed lines
show the statistical errors.
The g.s., $1^+$, and $3^+$ states of $^8$Be
have widths less than 250~keV and their $E(\tau)$ appear to be
converged and constant over the averaging regime.  The $2^+$ state
has a width of 1.5~MeV and the $4^+$ state's width is 3.5~MeV;
the $E(\tau)$ for the $2^+$ state might be converged, but that
of the $4^+$ state is clearly steadily decreasing.  Reliable
estimates of the energies of broad resonances requires the use
of scattering-state boundary conditions (see Sec.~\ref{sec:scat}) 
which have not yet been implemented for more than five nucleons.

Tests with different $\Psi_T$ as starting points for
the GFMC calculation are described in Ref.~\cite{PPCPW97}.
The most crucial aspect in choosing $\Psi_T$ is that it have the
correct mix of spatial symmetries, i.e., the $\beta_{LS[n]}$ admixtures.
This is because the limited propagation time is not sufficient for the GFMC
algorithm to filter out low-lying excitations with the same quantum numbers.
This issue is unimportant in $^4$He, where the first 0$^+$ excited state is
near 20 MeV, but in $^6$Li, the first 1$^+$ excited state is at only 5.65 MeV;
other $p$-shell nuclei have similar low-lying excitations.
Otherwise, the GFMC algorithm is able to rapidly correct for some very poor
$\Psi_T$ whose variational energies are actually positive.
However, a good $\Psi_T$ helps to keep the error bars small at larger $\tau$.
The most efficient balance between speed of construction of $\Psi_T$ and
smallest number of samples needed to achieve a given error bar seems to be
given by Eq.(\ref{eq:psitgfmc}).
A number of other tests of the GFMC algorithm and its ability to determine
nuclear radii are described in Refs.~\cite{PPCPW97,WPCP00}.

The calculations described here are computationally intense, and would not
have been possible without the advent of parallel supercomputers.
The initial studies~\cite{PPCW95} of $^6$Li required $\sim 2,000$ node hours
on an IBM SP1 to propagate 10,000 configurations to $\tau$ = 0.06 MeV$^{-1}$.
Today, with improvements in the algorithm and after much effort to optimize
the computer codes, the same calculation requires about 15 node hours on a
third generation IBM SP.
However, a present calculation for $^8$Li, requiring 10,000 configurations
to get a reasonable error bar, takes 1,000 node hours, so forefront
computer resources remain essential for this program.

\section{ENERGY RESULTS}

There are a number of accurate many-body methods for evaluating the binding
energy of the s-shell nuclei, $^3$H, $^3$He, and $^4$He, using realistic nuclear
forces.  These include Faddeev in configuration space~\cite{FPSS93} (FadR), 
Faddeev and Faddeev-Yakubovsky in momentum space~\cite{NKG00} (FadQ), 
and hyperspherical harmonics~\cite{HH} (HH), 
correlated hyperspherical harmonics~\cite{VKR95} (CHH)
and pair-correlated hyperspherical harmonics~\cite{KVR93} (PHH).
Some results of these methods for the AV18 and AV18/UIX Hamiltonians 
are shown in Table~\ref{tab:s-shell}.
We observe that the VMC upper bounds are generally 1.5--2.0\% above the GFMC
energies, which in turn are 0.25--1.0\% above the Faddeev results.
The disagreement between GFMC and the other methods may
be attributable to the fact that GFMC is propagated with an $H'$ and a
small piece of the Hamiltonian is computed in perturbation.
At this fine level of comparison, one needs to worry about how features such
as $T=\case{3}{2}$ admixtures and $n-p$ mass differences in the trinucleon
ground state are treated in each calculation.

Table~\ref{tab:s-shell} also shows the necessity of including (at least) a 
three-nucleon potential in the Hamiltonian in order to reproduce the A=3,4 
experimental binding energies.
This is true for all the other modern $N\!N$ potentials, as shown in 
Refs.~\cite{FPSS93,NKG00}.
The local potentials AV18, Reid93, and Nijm II, give very similar results, 
while the slightly nonlocal Nijm I gives 1--3\% more binding, and the
more nonlocal CD-Bonn gives 5--8\% more binding, or about halfway between
the local potential values and experiment.
At present, no two- plus three-nucleon potential combination gives an
exact fit to both trinucleon and $^4$He energies, but several combinations, 
like AV18/UIX, come quite close.
It also appears that it may be easier to fit $^3$He and $^4$He simultaneously,
rather than $^3$H and $^4$He, perhaps because of the better data constraints
on the $pp$ interaction compared to the $nn$ interaction~\cite{NKG00}.
In principle, there should also be four-body forces, but their contribution
must be quite small, and it is impossible to unambiguously identify a need 
for such terms until a more thorough survey of possible three-nucleon
potentials is made and the discrepancies between the various many-body 
methods are resolved.

\begin{table}
\caption{Energies of $^3$H and $^4$He in MeV, computed by VMC and 
GFMC compared to momentum-space Faddeev~\cite{NKG00}, 
configuration-space Faddeev~\cite{FPSS93} and various hyperspherical harmonics
methods~\cite{kievsky}.
The HH column values are PHH for $^3$H; HH for $^4$He, 
AV18; and CHH for $^4$He, AV18/UIX}
\label{tab:s-shell}
\begin{tabular}{llrrrrrrr}
\toprule
 Hamiltonian & nucleus &   VMC     &   GFMC    & FadQ     & FadR     & HH     & Expt. \\
\colrule
 AV18        & $^3$H   & --7.50(1) & --7.61(1) & --7.623  &--7.62    &--7.618 &  8.482 \\
             & $^4$He  &--23.72(3) &--24.07(4) &--24.28   &          &--24.18 & 28.296\\
\colrule                                                            
 AV18/UIX    & $^3$H   & --8.32(1) & --8.46(1) & --8.478  &          &--8.475 &  8.482 \\
             & $^4$He  &--27.78(3) &--28.33(2) &--28.50   &          &--28.1  & 28.296 \\
\botrule
\end{tabular}
\end{table}

\begin{figure}
\centerline{\psfig{figure=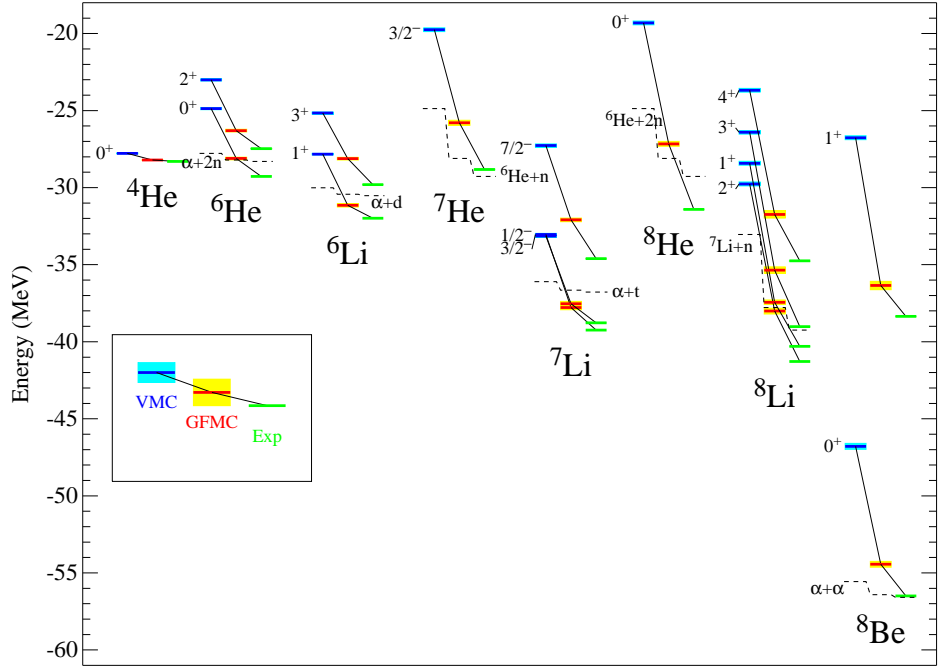,width=5.0in,angle=270}}
\caption{VMC and GFMC energies using AV18/UIX compared to experiment.
Black dashed lines show the indicated breakup thresholds for each method.
The Monte Carlo statistical errors are shown by the light blue and yellow bands.}
\label{fig:vmc-gfmc-exp}
\end{figure}

Figure \ref{fig:vmc-gfmc-exp} compares VMC (using the full $\Psi_V$ of
Eq.(\ref{eq:bestpsiv})) and GFMC calculations for the AV18/UIX
Hamiltonian, and also shows experimental energies.
All GFMC energies in this review are from Ref.~\cite{PPWC01};
VMC energies are from Refs.~\cite{WPCP00,PPCPW97}.
All particle-stable or narrow-width ($\Gamma<150$~keV)
states for $4\leq{A}\leq8$ are shown, except isobaric analogs.
We see that the VMC calculations get progressively worse 
in the p-shell compared to the s-shell, being on the order of 10--15\% above
the final GFMC results, for nuclei with $N \sim Z$.
In absolute terms, $\Psi_V$ misses roughly an extra 1.5 MeV of binding for
each p-shell nucleon that is added.
The VMC results also fail to reproduce important qualitative features
of the GFMC calculations; for example $^6$Li is stable against
breakup into $\alpha$+d for this Hamiltonian but the VMC calculation
shows it unbound by $\sim$2~MeV
(the black dashed lines show the 
indicated thresholds computed using the sub-cluster energies appropriate
to each calculation or experiment).
Clearly, there is some significant feature of light p-shell nuclei that 
is not yet included in the trial wave functions.

Table \ref{tab:energy} shows the GFMC energy values for several
Hamiltonians.  The table and Fig.~\ref{fig:vmc-gfmc-exp} show that the
AV18/UIX Hamiltonian, which gives excellent energies, compared to
experiment, for the s-shell nuclei, significantly underbinds the p-shell
nuclei.  The Li isotopes are stable against breakup into subclusters with AV18/UIX,
but progressively more underbound as $A$ increases.  There is also an
isospin problem, in that the He isotopes are even further off from the
experimental values, and in fact do not show stability against breakup
into subclusters.

\begin{table}
\caption{GFMC and experimental energies of $A$=3--8 ground states in MeV.
Energies of artificially confined 7- and 8-neutron drops are also shown.
The experimental values are from Refs.~\protect\cite{TUNL3,TUNL4,AS88}.}
\label{tab:energy}
\begin{tabular}{lcccl}
\toprule
                        & AV18         &  AV18/UIX    &  AV18/IL2    & ~~Expt. \\
\colrule
$^3$H($\case{1}{2}^+$)  & ~$-$7.61(1)  & ~$-$8.46(1)~ & ~$-$8.43(1)  & ~$-$8.48   \\
$^4$He(0$^+$)           & $-$24.07(4)~ & $-$28.33(2)~ & $-$28.37(3)~ & $-$28.30  \\
$^6$He(0$^+$)           & $-$23.9(1)~~ & $-$28.1(1)~~ & $-$29.4(1)~~ & $-$29.27  \\
$^6$Li(1$^+$)           & $-$26.9(1)~~ & $-$31.1(1)~~ & $-$32.3(1)~~ & $-$31.99  \\
$^7$He($\case{3}{2}^-$) & $-$21.2(2)~~ & $-$25.8(2)~~ & $-$29.2(3)~~ & $-$28.82  \\
$^7$Li($\case{3}{2}^-$) & $-$31.6(1)~~ & $-$37.8(1)~~ & $-$39.6(2)~~ & $-$39.24  \\
$^8$He(0$^+$)           & $-$21.6(2)~~ & $-$27.2(2)~~ & $-$31.3(3)~~ & $-$31.41  \\
$^8$Li(2$^+$)           & $-$31.8(3)~~ & $-$38.0(2)~~ & $-$42.2(2)~~ & $-$41.28  \\
$^8$Be(0$^+$)           & $-$45.6(3)~~ & $-$54.4(2)~~ & $-$56.6(4)~~ & $-$56.50  \\
\colrule
$^7$n($\case{1}{2}^-$)  & $-$33.47(5)  & $-$33.2(1)~  & $-$35.8(2)~  &          \\
$^8$n(0$^+$)            & $-$39.21(8)  & $-$37.8(1)~  & $-$41.1(3)~  &          \\
\botrule
\end{tabular}
\end{table}

\begin{figure}
\centerline{\psfig{figure=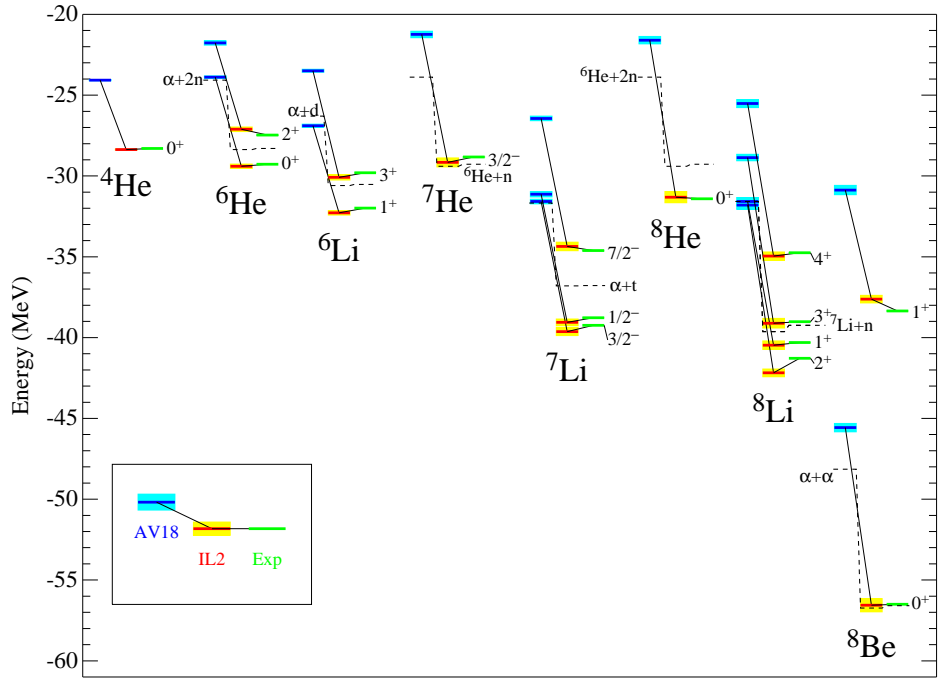,width=5.0in,angle=270}}
\caption{GFMC energies using AV18 and AV18/IL2 compared to experiment.}
\label{fig:av18-il2-exp}
\end{figure}

The new Illinois three-nucleon potentials were constructed to solve
the p-shell binding problems.  Figure~\ref{fig:av18-il2-exp} 
and Table~\ref{tab:energy} compare
GFMC calculations for AV18 with no $V_{ijk}$ and AV18/IL2 to 
the same experimental energies shown in Fig.~\ref{fig:vmc-gfmc-exp}.
The AV18/IL2 Hamiltonian does a good job of reproducing these energies;
the rms deviation from experiment for these levels is only
360~keV, while it is 2.3~MeV for AV18/UIX.  The AV18 values with
no $V_{ijk}$ show the large contribution that three-nucleon potentials
make to these binding energies; for $A=8$ the IL2 increases the
binding energy by more than 10~MeV.

Table~\ref{tab:sosplit} shows some splittings for states that in
a simple shell-model picture have the same $L$ and $S$ but different total $J$.
In all cases the lowest state for each $J$ is used.  Some of
the states, particularly the $^5$He states,  have large experimental widths;
calculations such as those described in Sec.~\ref{sec:scat} would be more
reliable.
The splittings are computed for the
AV18, AV18/UIX, and AV18/IL2 Hamiltonians and are compared to experimental
values.  The AV18 with no $V_{ijk}$ significantly underpredicts all the
splittings except the $\case{1}{2}^- - \case{3}{2}^-$ doublet in $^7$Li;
however in this case the relatively large statistical error for the small
splitting makes any conclusion difficult (the computed splittings are
the difference of two independent GFMC calculations whose statistical errors
must be added in quadrature).
The AV18/UIX Hamiltonian substantially improves the $^5$He splitting, and a 
VMC study showed that an earlier Urbana $V_{ijk}$  significantly increased the
computed splitting in $^{15}$N, resulting in much better agreement with
experiment~\cite{PP93}.  However the other splittings in the table are
still underpredicted with AV18/UIX.  The AV18/IL2 results in good predictions
of all the splittings except for the 1$^+-2^+$ doublet in $^8$Li which
is overpredicted.

\begin{table}
\caption{GFMC and experimental~\protect\cite{AS88,TUNL567} splittings of 
states with the same $L$ and $S$ but different $J$, in MeV.}
\label{tab:sosplit}
\begin{tabular}{lccccccc}
\toprule
       &                                  &$L$&  $S$        &  AV18  & AV18/UIX & AV18/IL2 & Expt.\\
\colrule
$^5$He &  $\case{1}{2}^- - \case{3}{2}^-$ & 1 & $\case{1}{2}$ & 0.6(1)~ & 1.1(2) & 1.3(2) & 1.20 \\
$^6$Li &  2$^+ - 3^+$                     & 2 &     1         & 0.8(1)~ & 0.9(1) & 2.2(2) & 2.12 \\
$^7$Li &  $\case{1}{2}^- - \case{3}{2}^-$ & 1 & $\case{1}{2}$ & 0.5(2)~ & 0.3(2) & 0.6(3) & 0.47 \\
$^7$Li &  $\case{5}{2}^- - \case{7}{2}^-$ & 3 & $\case{1}{2}$ & 0.7(2)~ & 0.8(2) & 2.2(3) & 2.05 \\
$^8$Li &  1$^+ - 2^+$                     & 1 &     1         & 0.2(4)~ & 0.6(2) & 1.7(4) & 0.98 \\
\colrule
$^7$n  &  $\case{3}{2}^- - \case{1}{2}^-$ & 1 & $\case{1}{2}$ & 1.65(7) & 1.5(1) & 2.8(3) &      \\
\botrule
\end{tabular}
\end{table}

A detailed breakdown of the GFMC ground-state energies for AV18/IL2
into kinetic, two- and
three-nucleon interactions is given in Table~\ref{tab:gfmc}.
Because of the extrapolation of the mixed expectation values,
Eq.(\ref{eq:pc_gfmc}), these components are not as accurate
as the total energy, and do not add up to the full amount;
indeed the sum of the individual kinetic and potential energies differs
from $\langle H \rangle$ by the same amount that the GFMC has improved
the $\Psi_T$ energy.
Nevertheless, they give a good idea about the relative size of the different
terms involved.
There is a big cancellation between the kinetic and two-body potential terms.
Consequently, while $V_{ijk}$ is less than 8\% of $v_{ij}$ in magnitude,
its expectation value is up to 50\% of the net binding (however
the net effect of the IL2 $V_{ijk}$, defined as the difference of the
binding energies computed without and with  $V_{ijk}$, is at most 30\%,
as can be seen in Table~\ref{tab:energy}).
This difference between expectation value and net effect is due to the large 
change that $V_{ijk}$ induces in $\langle K+v_{ij} \rangle$.

\begin{table}
\caption{Kinetic, $K$, and potential energy contributions in MeV. Shown are
the total NN and NNN potential energies, $v_{ij}$ and $V_{ijk}$;
the electromagnetic, $v^{\gamma}_{ij}$, and one-pion, $v^{\pi}_{ij}$, parts of 
$v_{ij}$; and the two- and three-pion parts of $V_{ijk}$.}
\label{tab:gfmc}
\begin{tabular}{lccccccl}
\toprule
       & $K$      & $v_{ij}$    & $V_{ijk}$ & $v^{\gamma}_{ij}$ 
                                                    & $v^{\pi}_{ij}$ & $V^{2\pi}_{ijk}$ 
                                                                             & ~~~$V^{3\pi}_{ijk}$ \\
\colrule                                            
$^3$H  & ~51.~~~~ & ~$-$59.~~~~ & ~$-$1.5~~~ & 0.04 & ~$-$45.~~~~ & ~$-$3.0~~~ &  ~~0.18(1) \\
$^4$He & 115.(1)  & $-$136.(1)  & ~$-$8.4(1) & 0.86 & $-$105.~~~~ & $-$16.3(1) &  ~~0.63(1)  \\
$^6$He & 147.(2)  & $-$171.(2)  & $-$11.5(3) & 0.87 & $-$127.(1)  & $-$20.3(4) & $-$0.91(6) \\
$^6$Li & 160.(2)  & $-$187.(2)  & $-$11.1(3) & 1.73 & $-$150.(1)  & $-$19.8(4) & $-$0.44(5) \\
$^7$He & 175.(3)  & $-$199.(3)  & $-$16.3(4) & 0.89 & $-$145.(2)  & $-$25.(1)~ & $-$3.1(1) \\
$^7$Li & 199.(3)  & $-$232.(3)  & $-$14.5(4) & 1.80 & $-$178.(2)  & $-$25.6(6) & $-$1.1(1) \\
$^8$He & 190.(3)  & $-$218.(3)  & $-$16.3(5) & 0.89 & $-$153.(1)  & $-$25.6(6) & $-$4.0(1) \\
$^8$Li & 242.(2)  & $-$278.(2)  & $-$20.6(4) & 1.93 & $-$211.(1)  & $-$34.2(5) & $-$3.8(1) \\
$^8$Be & 256.(4)  & $-$303.(3)  & $-$21.(1)~ & 3.32 & $-$234.(2)  & $-$38.5(9) & $-$0.9(2) \\
\colrule                                                                                     
$^7$n  & 105.(1)  & ~$-$59.(1)  & ~$-$3.6(3) & 0.07 & ~$-$10.~~~~ & ~$-$0.1(1) & $-$5.4(3) \\
$^8$n  & 122.(1)  & ~$-$73.(1)  & ~$-$3.0(3) & 0.09 & ~$-$12.~~~~ &   ~~0.3(1) & $-$5.9(4) \\
\botrule
\end{tabular}
\end{table}

Among the subcomponents of $v_{ij}$, the one-pion exchange dominates,
being 70--80\% of the total $v_{ij}$ for $A\geq3$ nuclei.
Similarly, the two-pion exchange is the dominant component of $V_{ijk}$.
The three-pion exchange term is small and repulsive in s-shell nuclei,
but attractive in the p-shell nuclei, which have $T=\case{3}{2}$ triples.
It is this term which results in the large improvement of the Illinois
models over the Urbana models for p-shell nuclei; however its contributions
are always less than 15\% of the two-pion $V_{ijk}$.
Finally, we note that the electromagnetic $v^\gamma_{ij}$ is dominated by the
Coulomb interaction between protons, but about 17\% (8\%) of
its total contribution comes from the magnetic moment and other terms
in He (Li) isotopes.

\begin{table}
\caption{GFMC isovector and isotensor energy coefficients, $a_n(A,T)$,
[Eq.~(\protect\ref{eq:isoexpand})] computed using AV18/IL2, in keV.
The contributions shown are kinetic energy, $K^{CSB}$; 
Coulomb potential, $v_{C1}(pp)$; other electromagnetic, $v^{\gamma,R}$;
and nuclear potential, $v^{CSB}+v^{CD}$.
The experimental value for $a_2(8,1)$ is extracted using
16.80 MeV as the excitation of the pure $T=1$ isobaric analog state 
in $^8$Be~\protect\cite{B66}.}
\label{tab:analog}
\begin{tabular}{lcrccrr}
\toprule
$a_n(A,T)$         & $K^{CSB}$& $v_{C1}(pp)$ & $v^{\gamma,R}$                               
                                                  & $v^{CSB}+v^{CD}$ & Total~~~ & Expt.\\
\colrule              
$a_1(3,\case{1}{2})$ &   14  &     649(1)   &    29   &   ~64(0)~ &  757(1)~ &   764\\ 
$a_1(6,1)$           &   16  &    1091(5)   &    18   &   ~47(1)~ & 1172(6)~ &  1173 \\
$a_2(6,1)$           &       &     166(1)   &    19   &   107(13) & 293(13)  &   223 \\ 
$a_1(7,\case{1}{2})$ &   22  &    1447(6)   &    40   &   ~79(2)~ & 1588(7)~ &  1644 \\
$a_1(7,\case{3}{2})$ &   18  &    1337(6)   &    12   &   ~52(1)~ & 1420(8)~ &  1373 \\
$a_2(7,\case{3}{2})$ &       &     137(1)   &    ~7   &   ~36(6)~ &  180(7)~ &   175 \\
$a_1(8,1)$           &   23  &    1686(5)   &    24   &   ~76(1)~ & 1810(6)~ &  1770 \\
$a_2(8,1)$           &       &     141(1)   &    ~4   &  $-$3(8)~ & 143(8)~  &   128 \\ 
$a_1(8,2)$           &   18  &    1528(7)   &    17   &   ~59(1)~ & 1622(8)~ &  1659 \\
$a_2(8,2)$           &       &     136(1)   &    ~6   &   ~38(5)~ & 180(5)~  &   153 \\ 
\botrule 
\end{tabular}
\end{table}

Energy differences among isobaric analog states are probes of the
charge-indepen-dence-breaking parts of the Hamiltonian.
The understanding of these ``Nolen-Schiffer energies'' has been
a theoretical problem for 30 years~\cite{NS69}.
The energies for a given isomultiplet of states can be expanded as
\begin{equation}
   E_{A,T}(T_z) = \sum_{n\leq 2T} a_n(A,T) Q_n(T,T_z) \ ,
\label{eq:isoexpand}
\end{equation}
where $Q_0=1$, $Q_1=T_z$, and $Q_2=\case{1}{2}(3T_z^2-T^2)$ are
isoscalar, isovector, and isotensor terms~\cite{P60}.
The isovector and isotensor coefficients $a_n(A,T)$, and various
contributions to them, are given in Table~\ref{tab:analog}.  
These were calculated as
expectation values in the $T_z=-T$ GFMC wave functions for AV18/IL2.
The individual terms are: 1) the effect of the neutron-proton mass
difference on the kinetic energy, $K^{CSB}$; 2) the proton-proton
Coulomb potential, including the AV18 form factor, $v_{C1}(pp)$; 
3) all other electromagnetic terms such as vacuum polarization
and magnetic-moment terms, $v^{\gamma,R}$; and
4) the strong-interaction contributions, $v^{CSB}$ which contributes
to the isotensor coefficients and $v^{CD}$ which contributes to
the isovector coefficients.

The $^3$H--$^3$He mass difference is 757 keV with the AV18/IL2 Hamiltonian,
in excellent agreement with the experimental value of 764 keV.
The bulk of the difference is the $pp$ Coulomb energy, but 108 keV
comes from the other terms.
Because of its very long range, the $v_{C1}(pp)$ dominates the $a_1(A,T)$
of heavier nuclei even more and small errors in the rms radius of the
nucleus can result in changes in the $v_{C1}(pp)$ contribution that
can totally mask the effects of the other, much smaller, terms.

\begin{table}
\caption{VMC isospin-mixing matrix elements for $^8$Be in keV computed
using AV18/UIX.  Experimental values are from Ref.~\protect\cite{B66}}
\begin{tabular}{lcccccc}
\toprule
$J^{\pi}$ & $K^{CSB}$& $v_{C1}(pp)$ & $v^{\gamma,R}$ & $v^{CSB}$  & $E_{01}$ & Expt. \\
\colrule
2$^+$     &       2  &      62    &      19  &      26    &  109(4)  &   149  \\
1$^+$     &       1  &      39    &       0  &      15    &  ~55(2)  &   120  \\
3$^+$     &       1  &      33    &      15  &      13    &  ~62(2)  &   ~63  \\
\botrule
\end{tabular}
\label{table:mixing}
\end{table}

Two (2$^+$;0+1) states occur very close together in the spectrum of $^8$Be at
16.6 and 16.9 MeV excitation; these isospin-mixed states come from blending the
(2$^+$;1) isobaric analog of the $^8$Li ground state with the second (2$^+$;0) 
excited state.
There are also fairly close (1$^+$;0,1) and (3$^+$;0,1) pairs at slightly
higher energies in the $^8$Be spectrum.
The isospin-mixing matrix elements that connect these pairs of states,
\begin{equation}
   E_{01}(J) = \langle \Psi(J^+;0) | H | \Psi(J^+;1) \rangle \ ,
\end{equation}
have been computed using VMC wave functions for the AV18/UIX Hamiltonian.
Results for the $E_{01}(J)$ are given in Table~\ref{table:mixing}.
The experimental values are determined from the observed decay widths
and energies~\cite{B66}.
The dominant contribution, from the Coulomb potential, typically 
accounts for less than half of the matrix element.
We see that the remaining part of the electromagnetic interaction and 
the strong CSB interaction can provide a significant boost, although
the experimental mixing is still underpredicted by $\sim$20\%.
It appears that these mixing elements are a more sensitive test
of the small CSB components of the Hamiltonian than are the isobaric analog
energy differences.

\section{DENSITY AND MOMENTUM DISTRIBUTIONS}

The one- and two-nucleon distributions of light $p$-shell nuclei
are interesting in a variety of experimental settings.
For example, the Borromean $^6$He and $^8$He nuclei are popular candidates 
for study as `halo' nuclei whose last neutrons are weakly bound.
In addition, the polarization densities of $^6$Li and $^7$Li are important
because of possible applications in polarized targets.

\begin{table}
\caption{GFMC values for point proton rms radii (in fm)
and quadrupole moments (in fm$^2$) in impulse approximation,
computed with AV18/IL2. 
Experimental values are from Refs.~\protect\cite{rmsr-expt,rmsr-3He,TUNL567,moments-A8}}
\label{tab:radii}
\begin{tabular}{cllll}
\toprule
   &\multicolumn{2}{c}{$\langle r^2_p \rangle^{1/2}$}
                                  &\multicolumn{2}{c}{$Q$} \\
          &     GFMC  & Expt.   & GFMC      &  Expt. \\
\colrule
$^3$H     &  1.59(0)  &  1.60   &             &             \\
$^3$He    &  1.76(0)  &  1.77   &           &             \\
$^4$He    &  1.45(0)  &  1.47   &             &             \\
$^6$He    &  1.91(1)  &         &             &             \\
$^6$Li    &  2.39(1)  &  2.43   &  --0.32(6)  &  --0.083   \\
$^7$Li    &  2.25(1)  &  2.27   &  --3.6(1)   &  --4.06     \\
$^7$Be    &  2.44(1)  &         &  --6.1(1)   &             \\
$^8$He    &  1.88(1)  &         &             &             \\
$^8$Li    &  2.09(1)  &         & \ms3.2(1)   & \ms3.11(5)  \\
$^8$B     &  2.45(1)  &         & \ms6.4(1)   & \ms6.8(2) \\
\botrule
\end{tabular}
\end{table}

Point proton rms radii and quadrupole moments are shown in 
Table~\ref{tab:radii}, as computed by GFMC for the AV18/IL2 model, along
with experimental values. 
The experimental charge radii have been converted 
to point proton radii by removing the proton and neutron $\langle r^2 \rangle$
of 0.743 and --0.116 fm$^2$, respectively.  
The radii are in good agreement with experiment, thanks at least in part
to the fact that the AV18/IL2 model reproduces the binding energies of these
nuclei very well.
The quadrupole moments have been calculated in impulse approximation;
two-body charge contributions are expected to provide only a few percent correction.
The agreement with experiment is again fairly good, with the exception 
of the $^6$Li quadrupole moment, which involves a delicate cancellation 
between the contributions from the deuteron quadrupole moment and the 
D-wave part of the $\alpha$+$d$ relative wave function.
In general, quadrupole moments are difficult to calculate accurately 
with quantum Monte Carlo methods because they are dominated by the long-range 
parts of the wave functions, which contribute very little to the 
total energy that VMC and GFMC are both designed to optimize.

\begin{table}
\caption{GFMC values for isoscalar and isovector magnetic moments
(in $\mu_N$) in impulse approximation using AV18/IL2.
Experimental values are from Refs.~\protect\cite{TUNL3,TUNL4,AS88,moments-A8}}
\label{tab:mag-mom}
\begin{tabular}{ccrrrr}
\toprule
            &    T      &\multicolumn{2}{c}{Isoscalar} & \multicolumn{2}{c}{Isovector} \\
            &                 &    GFMC &     Expt.   & GFMC     & Expt.  \\
\colrule                                              
$^3$He$-^3$H  & $\case{1}{2}$ &  0.403(0) &   0.426    & $-$4.330(1) & $-$5.107 \\
$^6$Li        &    0          &  0.817(1) &   0.822                             \\
$^7$Be$-^7$Li & $\case{1}{2}$ &  0.894(1) &   0.929    & $-$3.93(1)~ & $-$4.654 \\
$^8$B$-^8$Li  &    1          &  1.276(1) &   1.345    &   0.369(9) & $-$0.309 \\
\botrule
\end{tabular}
\end{table}

Calculated and experimental isoscalar and isovector magnetic moments are 
shown in Table~\ref{tab:mag-mom}, as defined with the convention used in 
Eq.(\ref{eq:isoexpand}); thus the isovector values for the $T=\case{1}{2}$ 
cases are $-$2 times those often quoted in the literature.
(For the $A=8$, $T=1$ nuclei the experimental isoscalar and isovector moments 
are obtained from the sum and difference of the values for B and Li, since the 
magnetic moment of the $T=1$ $J^{\pi}=2^+$ state in Be is not measured.)
The moments were calculated as expectation values in the GFMC wave functions 
with $T_z=-T$ in impulse approximation.
These impulse isoscalar moments are quite close to experiment and it is
expected that two-body current contributions are small~\cite{SPR89}.
For isovector moments, however, there can be significant corrections from
meson-exchange contributions.
We note that the corrections needed for the $A$=7,8 isovector moments are 
the same sign and magnitude as for $A$=3, so an exchange-current model that 
fixes the s-shell~\cite{MRS98} is likely to work for the light p-shell 
nuclei also.

\begin{figure}
\centerline{\psfig{figure=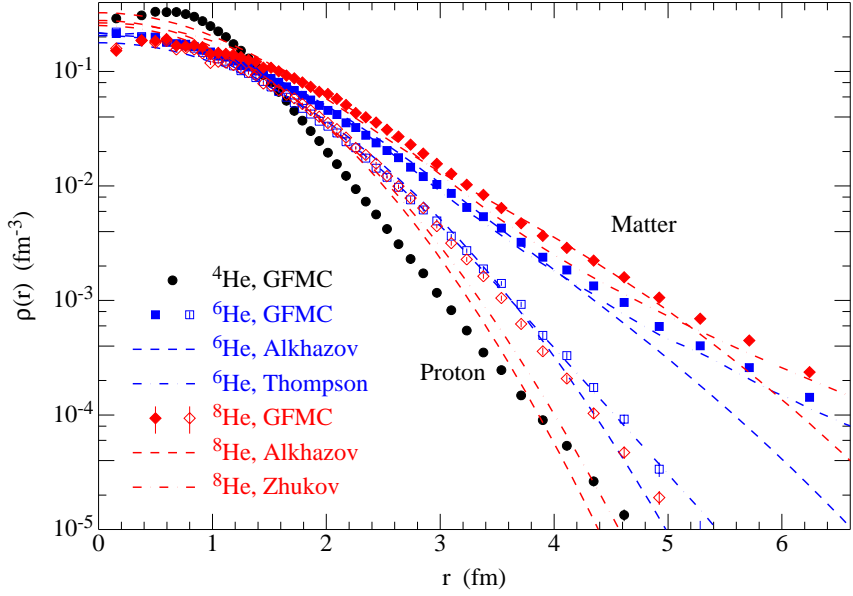,width=4.5in,angle=270}}
\caption{Proton and matter densities in $^4$He, $^6$He, and $^8$He.
Analyses of $^{6,8}$He-proton scattering are shown by the
dashed (Ref.~\protect\cite{Alkhazov}) 
and dot-dashed (Ref.~\protect\cite{Al-Khalili}) curves.}
\label{fig:he-rho1}
\end{figure}

In Fig.~\ref{fig:he-rho1}, we present proton and matter (proton$+$neutron) 
densities for the stable helium isotopes, as calculated with GFMC for AV18/IL2.
Nucleon densities are calculated as simple $\delta$-function expectation 
values, with possible spin and/or isospin projectors: for example, the
proton density is given by
\begin{equation}
\rho_{p}(r) \ = \ \frac{1}{4 \pi r^2}
\langle \Psi | \sum_i \ 
\ \frac { 1 + \tau_{zi}}{2} \ \delta ( r - | {\bf r}_i - {\bf R}_{cm} | )
\ | \Psi \rangle \ .
\end{equation}
The blue symbols and curves show results for $^6$He, red ones for $^8$He;
open symbols give the proton distributions which are also interpreted
as the alpha ``core'' density, full symbols are the total matter densities.
It can be seen that, as more neutrons are added, the tails of the matter 
distributions broaden considerably because of the relatively weak binding
of the p-shell neutrons.
In addition, the central neutron and proton densities decrease rather
dramatically.
This effect does not necessarily require any changes to the $\alpha$
core, but can be understood at least partially from the fact that the
$\alpha$ no longer sits at the center of mass of the entire system.
The motion relative to the center of mass spreads out the mass distribution
relative to that of $^4$He.  

The figure also shows two attempts to extract the densities of
$^{6,8}$He from scattering of beams of these short-lived nuclei from
proton targets~\cite{Alkhazov}.  The analysis of Alkhazov, et
al. (dashed curves) used Glauber theory and assumed proton (core) and
matter density distributions which were varied to fit the measured cross
sections.  The later work of Al-Khalili and Tostevin~\cite{Al-Khalili}
(dot-dashed curves) improved on this by doing the Glauber theory using
model wave functions which contained correlations between the 
valence neutrons.
The GFMC calculations definitely prefer the latter analysis.

\begin{figure}
\centerline{\psfig{figure=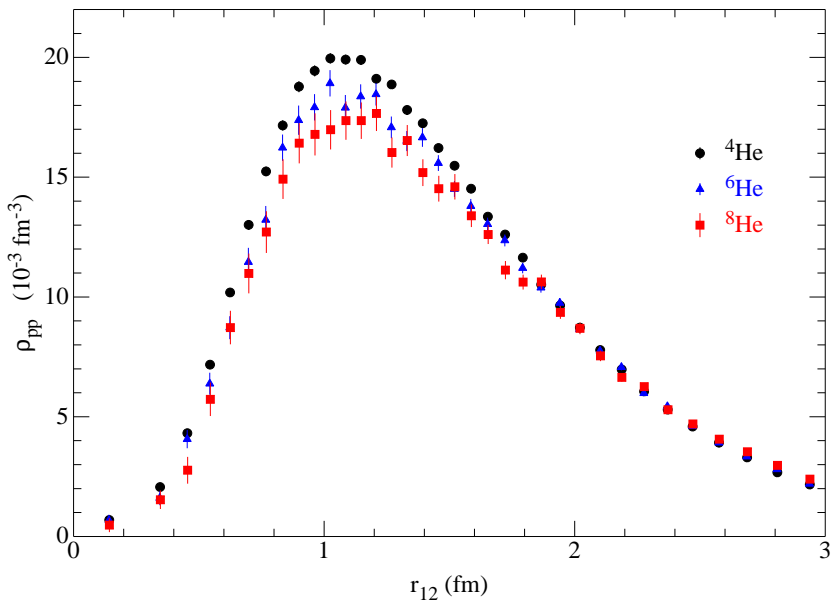,width=4.5in,angle=270}}
\caption{Proton-proton distribution functions in helium isotopes}
\label{fig:he-pp}
\end{figure}

The proton-proton distribution function is defined by
\begin{equation}
\rho_{pp} (r) = \frac{1}{4 \pi r^2}
\langle \Psi | \sum_{i<j} \ \frac{1 + \tau_{zi}}{2}
\ \frac { 1 + \tau_{zj}}{2} \ \delta ( r - |{\bf r}_i - {\bf r}_j|)|
\Psi \rangle \ .
\end{equation}
These distributions are directly related to the Coulomb sum measured in
inclusive longitudinal electron scattering; such measurements in $^3$He have
been used to put constraints on the $\rho_{pp} (r_{ij})$, and realistic
calculations agree with the experimental results\cite{SWC93}.
The behavior of $\rho_{pp} (r)$ at short distances is largely determined
by the repulsive core of the $N\!N$ potential and is nearly independent
of the nucleus, but at larger distances it is determined by the size
of the nucleus.

It is interesting to compare $\rho_{pp}$ for $^4$He to that of $^6$He and 
$^8$He.
These distribution functions are shown in Fig.~\ref{fig:he-pp}, again
calculated with GFMC for the AV18/IL2 model.
These nuclei each have just one $pp$ pair which presumably is in the
``alpha core'' of $^{6,8}$He.
Unlike the one-body densities, these distributions are not sensitive to
center of mass effects, and thus if the alpha core of $^{6,8}$He is
not distorted by the surrounding neutrons, all three $\rho_{pp}$ distributions
in the figure should be the same.
We see that the $pp$ distribution spreads out slightly with neutron
number in the helium isotopes, with an increase of the pair rms radius of
approximately 4\% in going from $^4$He to  $^6$He, and 8\% to $^8$He.
While this could be interpreted as a swelling of the alpha core, it might
also be due to the charge-exchange ($\tau_i \cdot \tau_j$) correlations
which can transfer charge from the core to the valence nucleons.
Since these correlations are rather long-ranged, they can have
a significant effect on the $pp$ distribution.
VMC calculations of $^4$He with wave functions modified to give
$\rho_{pp}$ distributions close to those of $^{6,8}$He suggest
that the alpha cores of $^{6,8}$He are excited by $\sim80$ and $\sim350$ keV,
respectively.

In general, VMC calculations give one-body densities very similar to the
GFMC results, although the two-body densities may differ by up to 10\% in
the peak, with the GFMC having a somewhat sharper structure.
Because the VMC wave functions are simpler and less expensive to 
construct they have been used in a number of applications
where the single-particle structure is dominant.
All these calculations were carried out with wave functions for the AV18/UIX 
Hamiltonian, but we expect that GFMC calculations with the improved AV18/IL2 
model will not qualitatively alter the results.
(Remember that the VMC wave functions are constrained to reproduce available
experimental rms radii.)

\begin{figure}
\centerline{\psfig{figure=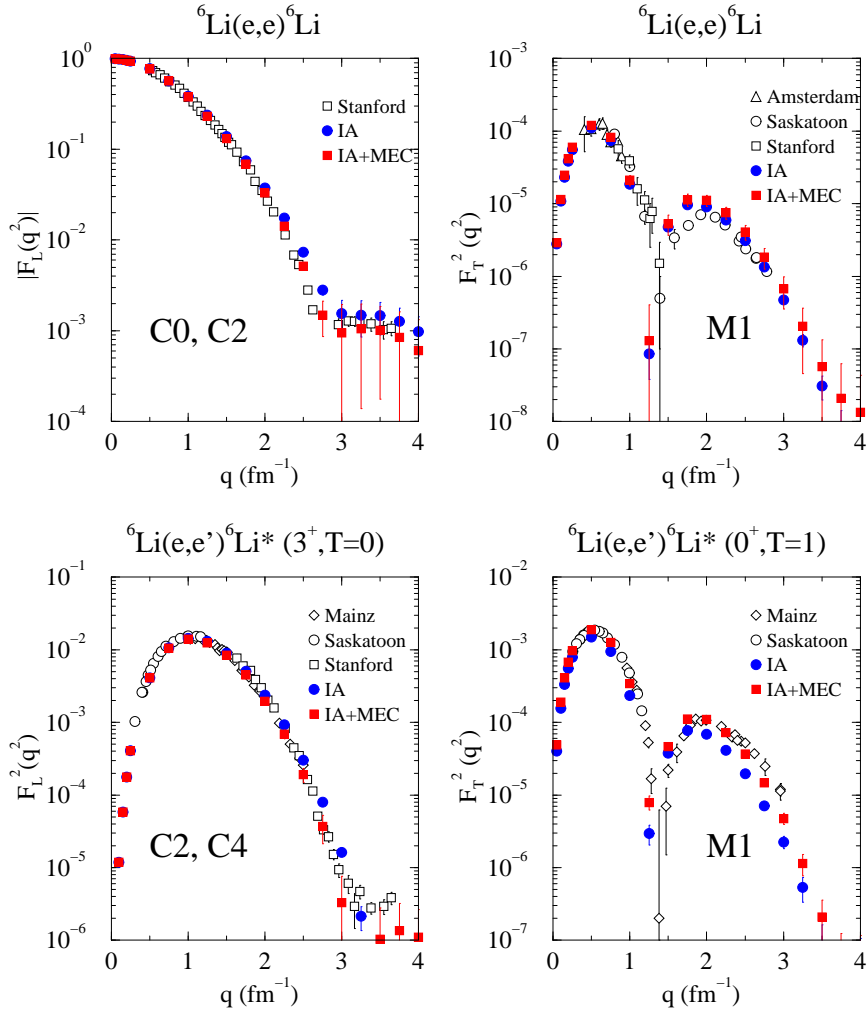,width=4.5in}}
\caption{Calculated and experimental elastic and transition form factors in 
$^6$Li.  Experimental data are from 
Refs.~\protect\cite{Li71,Lap78,Ber82,Ran66,Deu79,Ber76,Ber75}.}
\label{fig:li6ff}
\end{figure}

VMC calculations of elastic and inelastic electromagnetic form
factors for $^6$Li are shown in Fig.~\ref{fig:li6ff}.
These have been made in impulse approximation (IA) and with meson-exchange
contributions (MEC) to the charge and density operators~\cite{WS98}.
The comparison with data for the elastic longitudinal form factor, $F_L(q^2)$,
is excellent, as is the E2 transition to the 3$^+$ first excited state.
The MEC corrections are small, but stand out noticeably in the first minimum
where they significantly improve the fit to data.
The elastic transverse form factor, $F_T(q^2)$, is good up to the first zero,
but is a little too large in the region of the second maximum.
The M1 transition to the 0$^+$ isobaric analog state is also reproduced
reasonably well.
This kind of quantitative agreement with data has not been achieved in the
past with either shell model or $\alpha$+$d$ cluster wave functions.
In fact, shell model calculations normally require the introduction of 
effective charges, typically adding a charge of $\sim 0.5e$ to both the
proton and neutron, to obtain reasonable transition rates~\cite{PWK67}. 
No effective charges are used in the VMC calculations.

\begin{figure}
\centerline{\psfig{figure=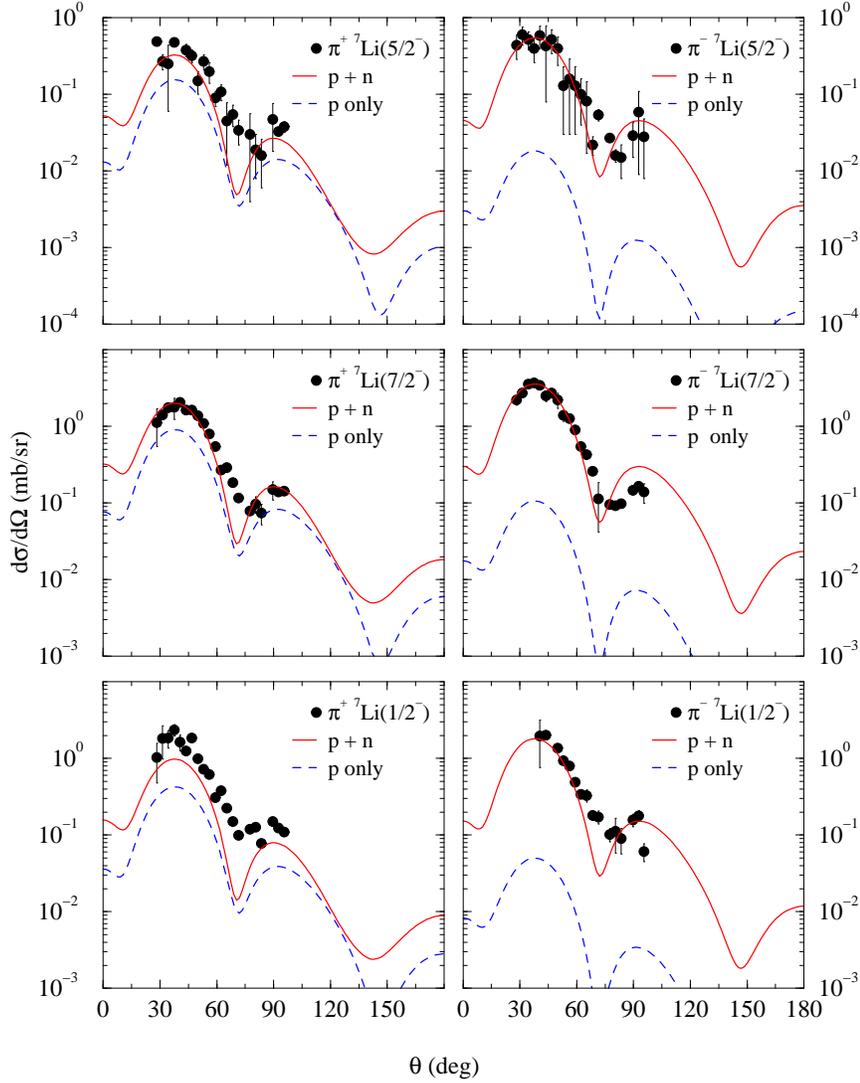,width=4.5in}}
\caption{Differential cross sections for $^7$Li$(\pi,\pi')$ scattering at
$E_\pi = 164$ MeV.  Data are from Ref.~\protect\cite{Bol79}.}
\label{fig:pipip}
\end{figure}

While electron scattering from nuclei is primarily sensitive to the proton
distributions in nuclei, $\pi^-$ scattering is most sensitive to the
neutron distributions.
Pion inelastic scattering at medium energy ($80 \leq E_{lab} \leq 300$ MeV) 
is dominated by strong absorption due to excitation of the $\Delta$ resonance,
and is well described in distorted-wave impulse approximation (DWIA).
An analysis of meson factory data for p-shell nuclei based on the Cohen-Kurath
shell model~\cite{CK65} showed that reasonable agreement with data 
could be achieved if the quadrupole excitation operator was enhanced 
by a factor $\sim$ 1.75--2.5~\cite{LK80}.
Figure~\ref{fig:pipip} shows a calculation of the differential cross sections
for $^7$Li($\pi,\pi^\prime$) scattering to the first three excited states
using VMC transition densities as input, with no enhancement 
factors~\cite{LW01}.
Solid red lines show the full results, which are in good agreement with the
data, while the dashed blue lines give the contribution coming from protons 
only.
Because of the strong isospin dependence of the $\pi N$ scattering $t$-matrix,
this reaction can be a stringent test of the neutron densities predicted
by the quantum Monte Carlo calculations.

\begin{figure}
\centerline{\psfig{figure=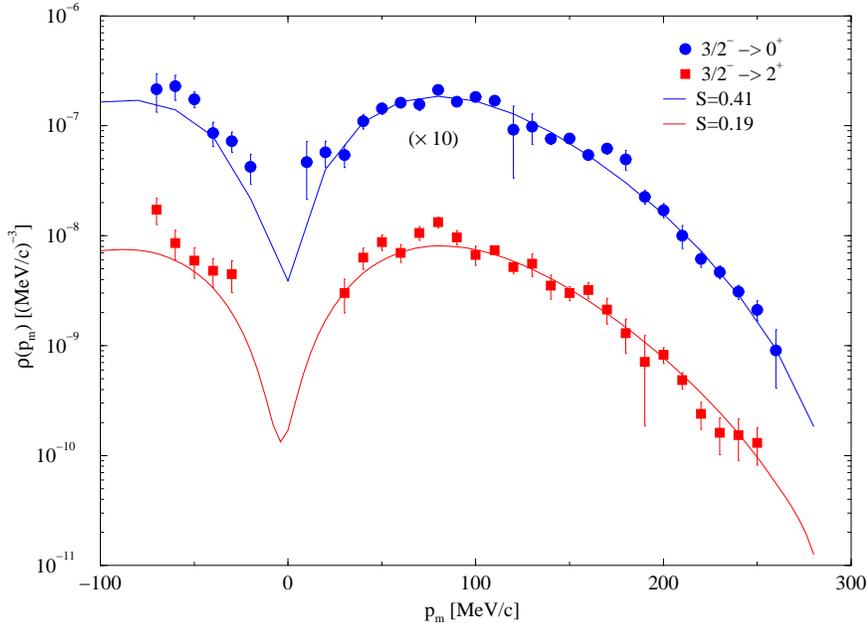,width=4.5in,angle=270}}
\caption{Experimental~\protect\cite{LWW99} momentum distributions for the 
transitions to the ground state (circles) and first excited state (squares) 
in the reaction $^7$Li$(e,e'p)^6$He, compared to CDWIA calculations with 
the VMC wave functions (solid lines) for AV18/UIX. }
\label{fig:he6p}
\end{figure}

The VMC wave functions for the AV18/UIX model have also been used to
calculate single-nucleon momentum distributions in many nuclei,
and a variety of cluster-cluster overlap wave functions, such as
$\langle d p | t \rangle$, $\langle d d | \alpha \rangle$, and $\langle
\alpha d | ^6{\rm Li} \rangle$~\cite{SPW86,FPPWSA96}.
Recently the overlaps 
$\langle^6{\rm He}(J^{\pi})+p(\ell_j)|^7{\rm Li}\rangle$ for 
all possible p-shell states in $^6$He were studied~\cite{LWW99}.
The spectroscopic factors obtained from these overlaps are 0.41 to the 
ground state of $^6$He and 0.19 to the 2$^+$ first excited state.
These factors are significantly smaller than the predictions of the 
Cohen-Kurath shell model~\cite{CK67}, which gives values of 0.59 and 0.40, 
respectively.
The CK shell model requires that the possible $^6$He$+p$ states sum to unity
within the p-shell, whereas in the VMC calculation, the correlations in 
the wave function push significant strength to higher momenta that cannot 
be represented as a $^6$He state plus p-wave proton.
The VMC overlaps were used as input to a Coulomb DWIA analysis of recent 
$^7$Li$(e,e^{\prime}p)^6$He data taken at NIKHEF~\cite{LWW99} and found to 
give an excellent fit to the data, as shown in Fig.~\ref{fig:he6p}.

\begin{figure}
\centerline{\psfig{figure=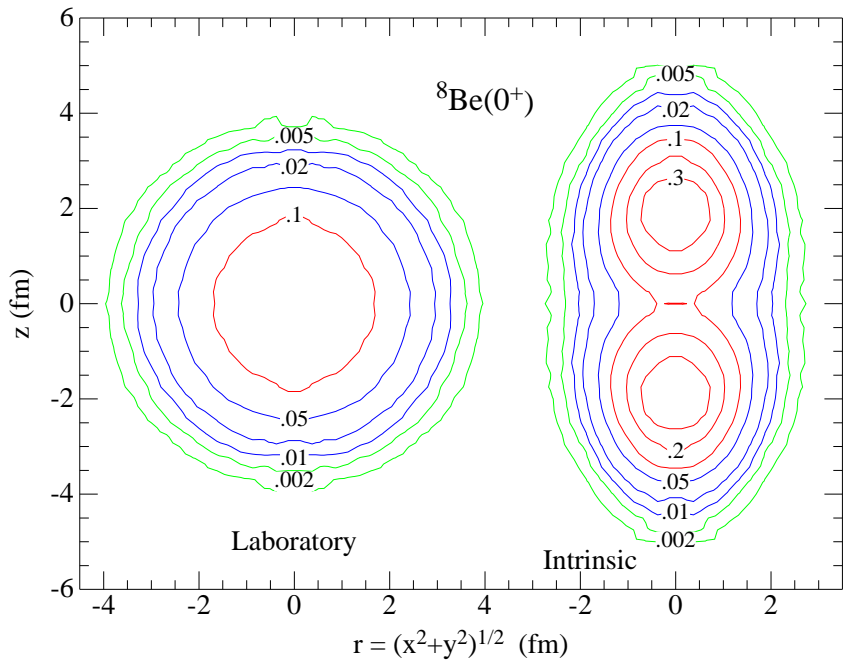,width=4.5in}}
\caption{Calculated density contours of $^8$Be in the lab frame (left) and the
intrinsic frame (right), labeled with densities in fm$^{-3}$}
\label{fig:be8}
\end{figure}

The VMC wave functions are based on one-body parts that have
a shell-model structure, namely four nucleons in an $\alpha$ core coupled to
$(A-4)$ one-body $\ell=1$ wave functions.  However the low-lying
states of $^8$Be exhibit a rotational spectrum and are believed
to be well approximated as
two $\alpha$'s rotating around their common center of mass.
It is possible to recover this picture from the VMC wave functions by a 
modified Monte Carlo density calculation~\cite{WPCP00}.

The standard Monte Carlo method for computing one-body densities, 
$\rho({\bf r})$, is to make a random walk that samples 
$|\Psi({\bf r}_1,{\bf r}_2,\cdots,{\bf r}_A)|^2$ and 
to bin ${\bf r}_1,{\bf r}_2,\cdots,{\bf r}_A$ 
for each configuration
in the walk.  The density is then proportional to the number
of samples in each bin.  In the case of a $J$~=~0 nucleus, this
``laboratory'' density will necessarily be spherically symmetric.

The intrinsic density in body-fixed coordinates can be approximated 
by computing the moment of inertia matrix, ${\cal M}$, of the $A$ positions
for each configuration.
The eigenvalues and eigenvectors of ${\cal M}$, are found and a
rotation to those principal axes is made.  The resulting 
${\bf r}_1^\prime,{\bf r}_2^\prime,\cdots,{\bf r}_A^\prime$ is then binned.
The eigenvector with the largest eigenvalue is chosen as the
${\bf z}^\prime$ axis.
This procedure will not produce a spherically symmetric distribution,
even if there is no underlying deformed structure, because 
almost every random configuration will have  principal axes of different
lengths and the rotation will always orient the longest principal axis
in the ${\bf z}^\prime$ direction.  However no 
artificial structure is introduced~\cite{WPCP00}.

When the above procedure is applied to the three lowest $^8$Be states,
a dramatic intrinsic structure is revealed, as shown in Fig.~\ref{fig:be8}
for the ground state.  
The figure shows contours of constant density plotted in cylindrical 
coordinates.
The left side of the figure shows the standard, lab frame, density 
calculation.
For the $J$~=~0 ground state, this is spherically symmetric as shown.
The right side of the figure shows the intrinsic density.
It is clear that the intrinsic density has two peaks, with the neck
between them having only one-third the peak density; we regard these
as two $\alpha$'s.  This assignment is strengthened by making
the same construction for the $J=2^+$ and $4^+$ states; although
the laboratory densities for these states (in $M=J$ states) are
quite different, the intrinsic densities are, within statistical errors,
the same as the $J=0$ intrinsic density~\cite{WPCP00}.

If the $0^+$, $2^+$, and $4^+$ states are generated by rotations
of a common deformed structure, then their electromagnetic moments
and transition strengths should all be related to the intrinsic
moments which can be computed by integrating over the
projected body-fixed densities.  This is explored in Ref.~\cite{WPCP00}
and a generally consistent picture is shown.

\section{LOW-ENERGY NUCLEON-NUCLEUS SCATTERING}
\label{sec:scat}

The calculations presented so far have treated resonant nuclear states
as if they were particle stable; that is the VMC trial wave functions
decay exponentially at large distances and the GFMC propagation
does not impose a scattering-state boundary condition.  As was
shown in Fig.~\ref{fig:8be-e_of_tau}, this GFMC propagation converges
for unstable states that are narrow, but for wide states, such as
$^8$Be(4$^+$), the $E(\tau)$ keeps decreasing for increasing $\tau$.
Such states should be computed using scattering-wave boundary conditions.

The $^5$He($\case{1}{2}^-$, $\case{3}{2}^-$) states have been studied
in VMC~\cite{csk87} and GFMC~\cite{cs94} using scattering-wave 
boundary conditions.  The VMC calculations use a trial function
that is a straight-forward generalization of the $\Psi_V$ given
in Sec. 3.  In it the $\Psi_J$ contains one p-wave single-particle function
that goes to zero at a specified (large) radius, $R_n$.  The variational
energy is minimized subject to this boundary condition.  The
phase shift at the resulting scattering energy, 
$E_s=E(^5\mbox{He}(R_n))-E(^4\mbox{He})$ is then obtained from
\begin{equation}
\tan(\delta_l) = \frac{j_l(k R_n)}{n_l(k R_n)},
\end{equation}
where $k = (2 \mu E_s)^{1/2} / \hbar$,
$\mu$ is the $\alpha$+$n$ reduced mass, and $j_l$ and $n_l$ are
the spherical Bessel functions (in this case $l=1$).  It is possible
to generalize this formulation by specifying a logarithmic-derivative
at $R_n$, rather than having the wave function go to zero there.
This method requires the computation of the energy difference, $E_s$.
If this is done with two different VMC calculations, one for $^5$He
with the specified boundary condition, and one for $^4$He, then
the statistical errors of these two calculations must be added
in quadrature.  Instead one can directly evaluate the energy
difference in a VMC random walk that is controlled by the
$^5$He wave function.  In practice this gives $E_s$ with a smaller
statistical error than either of the separate statistical errors~\cite{csk87}.
The GFMC calculations start with the scattering-state trial
function and use a modified propagator that preserves the
boundary condition~\cite{cs94}.  Both methods result in a phase shift
and energy corresponding to a given boundary condition.  By repeating
the calculations for different boundary conditions, the phase shift
as a function of energy can be mapped out.

The VMC calculations of Ref.~\cite{csk87} used older two- and three-nucleon
potentials than have been used for the other calculations reported
in this review.  They reproduced the qualitative features of the 
experimental $\alpha$+$n$ $\case{1}{2}^-$ and $\case{3}{2}^-$ phase shifts,
but, in particular, gave a spin-orbit splitting about 1~MeV too small.
The GFMC calculations~\cite{cs94} used a different, but still somewhat old
Hamiltonian and obtained values that are closer to experiment.  These
calculations need to be repeated with the modern Hamiltonians
described in this review, and for other systems such as $\alpha$+$d$,
$^6$He+$n$, and  $\alpha$+$\alpha$.

\section{ASTROPHYSICAL ELECTROWEAK REACTIONS}

Most of the key nuclear reactions in primordial nucleosynthesis~\cite{NB00}
and solar neutrino production~\cite{A+98} involve only the $A \leq 8$ nuclei.
Many of these are electroweak capture reactions that are difficult or
impossible to measure in the laboratory.
With the high-precision few-body methods, such as PHH and Faddeev, or new 
effective field theory treatments, the reactions to s-shell final states
have been calculated with unprecedented accuracy in the last few years.
These include the $^1$H$(p,e^+\nu_e)^2$H~\cite{SS+98} and 
$^3$He$(p,e^+\nu_e)^4$He~\cite{MSVKR00} weak capture reactions, and the
$^1$H$(n,\gamma)^2$H~\cite{R00}, $^2$H$(n,\gamma)^3$H and 
$^2$H$(p,\gamma)^3$He~\cite{VSK96} radiative capture reactions.

Recently, the VMC method has been applied to some of the radiative captures 
to p-shell final states, including $^2$H($\alpha,\gamma)^6$Li,
$^3$H($\alpha,\gamma)^7$Li, and $^3$He($\alpha,\gamma)^7$Be~\cite{NWS01,N01}.
As discussed in the previous section, many-nucleon scattering states can be
constructed within the quantum Monte Carlo framework, but it
has not yet been done for $A \geq 6$ nuclei.
Thus for these first studies, the appropriate electromagnetic
matrix elements (primarily E1 and E2) are evaluated between a VMC six- or
seven-nucleon final state, and a correlated cluster-cluster scattering 
state which is not variationally improved.

Since much of the capture takes place at long range, an important ingredient 
in these calculations is an asymptotically-correct description of the six- 
or seven-body final state.
This feature has to be implemented in the initial variational wave function 
because, as remarked before, the VMC and GFMC methods find their best wave
functions by optimizing the energy, which is not sensitive to the long-range
behavior.
The asymptotic behavior can be imposed by requiring the one-body wave 
function in Eq.(\ref{eq:phi}) to satisfy the condition:
\begin{equation}
\label{eqn:asymptotic}
[\phi^{LS[n]}_{p}(r\rightarrow\infty)]^n \propto W_{km}(2\gamma r)/r \ ,
\end{equation}
where $W_{km}(2\gamma r)$ is the Whittaker function for bound-state wave
functions in a Coulomb potential and $n$ is the number of p-shell nucleons.
Here $\gamma^2 = 2\mu_{4n} B_{4n}/\hbar^2$, with $\mu_{4n}$ and $B_{4n}$
the appropriate two-cluster effective mass and binding energy.
For $^6$Li, $B_{42} = 1.47$ MeV; for $^7$Li or $^7$Be, capture
can be to the $\case{3}{2}^-$ ground or $\case{1}{2}^-$ 
excited state with corresponding values for $B_{43}$.

The initial-state wave functions are taken as elastic-scattering
states of the form
\begin{equation}
\label{eqn:scatstate}
|\psi_{\alpha\tau}; LSJM \rangle =
{\cal A}\left\{\phi_{\alpha\tau}^{JL}(r_{\alpha\tau})Y_{LM_L}({\bf\hat{r}}_{\alpha\tau})
\prod_{ij}G_{ij}|\psi_\alpha\psi_\tau^{m_S}\rangle\right\}_{LSJM},
\end{equation}
where curly braces indicate angular momentum coupling, $\cal{A}$
antisymmetrizes between clusters, $\psi_\alpha$ is the $^4$He ground
state, and $\psi_\tau^{m_S}$ is the deuteron or trinucleon ground state in 
spin orientation $m_S$.
The $G_{ij}$ are identity operators if the nucleons $i$ and $j$ are in
the same cluster, else, they are a set of pair correlation operators, including
both central and spin-isospin dependent terms, which introduce distortions in
each cluster, under the influence of individual nucleons from the
other cluster.  They are similar to the correlations discussed in Sec.~3,
except that they revert to the identity operator at pair separations beyond 
about 2 fm.
The correlations $\phi_{\alpha\tau}^{JL}$ are generated from optical
potentials that describe the experimental phase shifts in cluster-cluster 
scattering; see Refs.~\cite{NWS01,N01} for details.
Because the $G_{ij}$ go to unity, the $\psi_{\alpha\tau}$ has the same
phase shifts as those generated by the optical potential.

Evaluation of electromagnetic matrix elements between the initial
scattering and final bound states can be split into two parts by noting 
that all of the energy dependence is contained in the relative wave 
function $\phi_{\alpha \tau}^{JL}$ and the transition operators.
Thus, using a technique first applied in a VMC calculation of $d$+$d$
radiative capture~\cite{APS91}, 
the matrix element for a given scattering partial wave can be written as
\begin{eqnarray}
\label{eqn:integrand}
T^{LSJ_iJ_f}_\ell(q) &=& \int_0^\infty dx\, x^2 \, \phi_{\alpha\tau}^{J_iL}(x)
 \\
&\times& \langle \psi_{A}^{J_fm_f}|
T_{\ell \lambda}(q) {\cal A}\left\{\delta(x-r_{\alpha\tau} )
Y_L^{M_L}({\bf\hat{r}}_{\alpha\tau}))
\prod_{ij}G_{ij}|\psi_\alpha\psi_\tau^{m_S}\rangle\right\}_{LSJ_iM_i} \ ,
\nonumber
\end{eqnarray}
where $T_{\ell \lambda}$ denotes any of the standard $E_{\ell}$ or 
$M_{\ell}$ operators.
The integration over all coordinates except $x$ can be calculated just once
for each partial wave by Monte Carlo sampling, and the result can then
be used to compute the full integral for as many energies as desired by
recomputing $\phi_{\alpha\tau}^{J_iL}(x)$ only.

\begin{figure}
\centerline{\psfig{figure=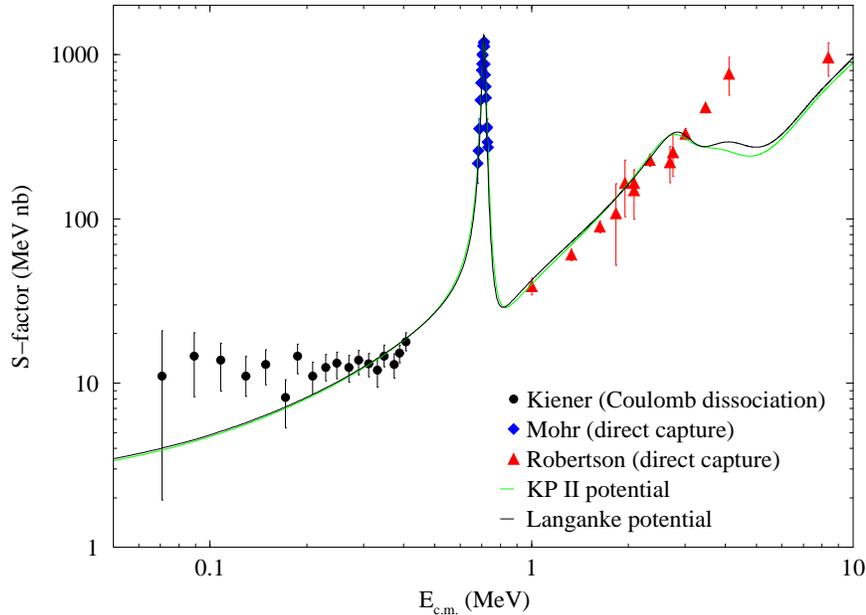,width=4.5in,angle=270}}
\caption{The total $S$-factor for $\alpha$+$d$ radiative capture calculated for
two different optical potentials with the AV18/UIX nuclear wave functions,
compared to experimental data~\protect\cite{rghr,mohr,kiener}.}
\label{fig:sfactor}
\end{figure}

The result for the $^2$H($\alpha,\gamma)^6$Li capture $S$-factor is shown
in Fig.~\ref{fig:sfactor}, where it is compared to a collection of direct
and indirect data.
The $S$-factor falls rapidly at low-energy because a pseudo-orthogonality
between the initial and final states suppresses normal S-wave capture.
Whenever there is such a suppression, it is important to consider higher-order
terms that might contribute, including relativistic corrections and two-body 
charge and current operators~\cite{CS98}.
This is possible with the VMC calculation because of the fully
correlated $A$-body wave functions that are used.
In this case, a relativistic center-of-energy correction leads to an E1 
contribution that dominates at low energy, while the resonance region and
above is primarily E2.
Results for $^3$H($\alpha,\gamma)^7$Li, and $^3$He($\alpha,\gamma)^7$Be
can be found in Ref.~\cite{N01}

This kind of VMC calculation is only a first investigation into the p-shell
astrophysical reactions, and should be improved in the future by using the
more exact GFMC wave functions for both bound and scattering states, and
the Hamiltonians with improved $N\!N\!N$ potentials.

\section{NEUTRON DROPS}

Neutron-rich nuclei are interesting both because of
their importance in various astrophysical contexts, such as
the r-process or neutron-star crusts, and the current 
interest in experiments with radioactive beams.  These nuclei are
often studied within the framework of mean-field models using
Skyrme or other potential models.  The parameters of these models
are fit to experimentally known binding energies, that is for
situations with $N{\sim}Z$.  In particular the 
isospin dependence of the spin-orbit component
of such potentials is considered to be not strongly constrained.
Neutron drops offer the possibility of theoretical guidance
for the isotopic dependence of such parameters.

Neutron drops are systems of interacting neutrons confined in an
artificial external well.  Eight neutrons form a closed shell
and single-particle spin-orbit splittings can be studied in
drops of seven or nine neutrons.  Pairing energies can be studied
if six-neutron drops are also computed.
Calculations of systems of seven and eight neutrons interacting with 
AV18/UIX were used as a basis for comparing Skyrme
models of neutron-rich systems with microscopic calculations based on
realistic interactions~\cite{PSCPPR96}.  
The external one-body well used is a Woods-Saxon:
\begin{equation}
V_1 (r) = \sum_i \frac{V_0}{1 + \exp [ - (r_i - r_0)/a_0 ]} \ ;
\end{equation}
the parameters are $V_0 = -20$ MeV, $r_0 = 3.0$ fm, and $a_0 = 0.65$ fm.
Neither the external well nor the total internal potential
($v_{ij}+V_{ijk}$) are individually attractive enough to produce bound
states of seven or eight neutrons; however the combination does produce
binding.

Tables \ref{tab:energy}-\ref{tab:gfmc} show results for the neutron drops
with this external well.  
The $T=\case{3}{2}$ nature of the
$S^I_{\tau}S^I_{\sigma}$ term of $V^{3\pi}$ results in large contributions
in the neutron drops.  As a result the seven-neutron drops computed with
AV18/IL2 have double the spin-orbit splitting predicted by AV18/UIX.
Thus the spin-orbit splitting in neutron-rich systems depends strongly on
the Hamiltonian used.

\section{CONCLUSIONS AND OUTLOOK}

Quantum Monte Carlo methods can now be used to obtain accurate (within 2\%)
energies for light p-shell nuclei up to $A$ = 8.
Initial calculations of $A$=9,10 nuclei have been made by the authors, and 
studies of $A$=11,12 nuclei should be feasible in the next few years by
variational and Green's function Monte Carlo methods.
This progress in the nuclear many-body problem is due both to the rapid 
growth of computational power and the continuing evolution of algorithms.
The new auxiliary-field diffusion Monte Carlo method, which samples spins and
isospins by auxiliary fields, and space by standard diffusion Monte Carlo,
may be the key to doing even larger nuclear systems~\cite{SF99}.

Studies of the p-shell nuclei allow us to test nuclear forces in new ways
not accessible in s-shell nuclei, particularly the odd partial waves of
$N\!N$ scattering and the $T=\case{3}{2}$ triples for $N\!N\!N$ forces.
They also give us many more cases in which to examine charge-dependent and 
charge-symmetry-breaking interactions.
With these calculations of nuclear spectra, we see for the first time that
nuclear structure, including both single-particle and clustering aspects,
really can be explained directly in terms of bare 
nuclear forces that fit $N\!N$ data.
A crucial ingredient for quantitative agreement is the addition of realistic 
$N\!N\!N$ forces, including at least two terms beyond the standard 
long-range two-pion-exchange potential~\cite{PPWC01}.

Many aspects of nuclear structure and reactions are described
quantitatively in these studies.
The energy differences between isobaric analog states are explained well once
the complete electromagnetic and strong charge-independence-breaking 
forces deduced from $N\!N$ scattering are included.
Charge radii and quadrupole moments agree with experiment, and we expect 
magnetic moments will also once the important two-body exchange currents 
are included, as they have been in s-shell nuclei~\cite{MRS98}.
Elastic and transition form factors measured in electron scattering,
and transition densities that are tested in pion scattering,
are accurately predicted without the introduction of effective charges.
Spectroscopic factors are naturally quenched by the correlations in
$A$-body wave functions and we see that those correlations build up
intrinsic cluster structure, as in the case of $^8$Be.

Initial calculations of nucleon-nucleus scattering states and electroweak
capture reactions are encouraging.
These two related problems will be a major activity for the quantum Monte
Carlo studies over the next several years, with studies of scattering states 
in nuclei like $^7$He and $^9$He, and of the cluster-cluster scattering
states that enter into the astrophysical capture reactions.
Additional reactions like $^7$Be$(p,\gamma)^8$B and even 
$^8$Be$(\alpha,\gamma)^{12}$C should become feasible.
Still other projects will include the study of weak decays and, going 
beyond the p-shell, looking at the unnatural-parity intruder states which 
start to dip into the low-energy spectrum with the $A=9$ nuclei, and can be
particle stable by $A=10$.

Refining the nuclear Hamiltonian will remain a major aspect of future work.
Extensive new $N\!N$ scattering data taken since the Nijmegen partial-wave 
analysis has increased the database to more than 6000 points, and a new 
CD-Bonn~2000 potential has been constructed that again achieves a 
$\chi^2/$datum $\approx 1$~\cite{M01}.
Also, intensive searches are being made to look for $N\!N\!N$ force signatures
in $N+d$ scattering experiments~\cite{K+01,W+01}, which may well indicate a
need for additional terms beyond those contained in the Illinois models.
Our ability to test new force models, as they become available, in light 
nuclei by accurate quantum Monte Carlo methods will continue to be an 
important tool for nuclear physics.

\section*{Acknowledgments}

The authors thank J. Carlson, A. Kievsky, V. R. Pandharipande,
R. Schiavilla, J. P. Schiffer and K. E. Schmidt for useful communications and
discussions.  This work was supported by the U.S. Department of Energy,
Nuclear Physics Division, under contract No. W-31-109-ENG-38.

\end{document}